\documentclass[11pt,a4paper]{article}
\pdfoutput=1
\usepackage{jheppub}
\usepackage[T1]{fontenc}
\usepackage{amsfonts}
\usepackage{amsmath}
\usepackage{color}
\usepackage[normalem]{ulem}
\usepackage{verbatim}

\usepackage{subcaption}

\newcommand{\beq}{\begin{equation}}
\newcommand{\eeq}{\end{equation}}
\newcommand{\bea}{\begin{eqnarray}}
\newcommand{\eea}{\end{eqnarray}}
\newcommand{\dd}{\text{d}}

\newcommand{\bra}[1]{\Bigl\langle #1\Bigr|}
\newcommand{\ket}[1]{\Bigl|#1\Bigr\rangle}
\newcommand{\ldot}{\!\cdot\!}

\newcommand{\ep}{\epsilon}

\newcommand{\wt}[1]{\widetilde{#1}}
\newcommand{\bd}[1]{\mathbf{#1}}
\newcommand{\wb}[1]{\overline{#1}}
\newcommand{\gtap}{\;\rlap{\raisebox{0.5ex}{$>$}}{\raisebox{-0.5ex}{$\sim$}}\;}

\title{Soft gluon evolution and non-global logarithms}
 \author{Ren\'e
\'Angeles Mart\'inez${}^a$, Matthew De Angelis${}^b$, \\Jeffrey R. Forshaw${}^b$,
Simon Pl\"atzer${}^{c}$ and Michael H. Seymour${}^{b,d}$}
\affiliation{${}^a$ The Henryk Niewodniczanski Institute of Nuclear Physics in
  Cracow,\\\hspace*{2ex}Polish Academy of Sciences PL-31342 Krakow,
  Poland\\ ${}^b$ Consortium for Fundamental Physics, School of Physics \&
  Astronomy,\\\hspace*{2ex}University of Manchester, Manchester M13 9PL, United
  Kingdom\\ ${}^c$ Particle Physics, Faculty of Physics,\\
  \hspace*{2ex}University of Vienna, 1090 Wien, Austria\\
  ${}^d$ Theoretical Physics Department,\\
  \hspace*{2ex}CERN, CH-1211 Geneva 23, Switzerland}

\preprint{\begin{flushright}MAN/HEP/2018/001\\CERN-TH-2018-037\\IFJPAN-IV-2018-4\\UWTHPH-2018-3
    \\ MCnet-18-03\end{flushright}}

\abstract{We consider soft-gluon evolution at the amplitude level. Our
  evolution includes Coulomb exchanges and applies to generic hard-scattering
  processes involving any number of coloured partons. We emphasise the special
  role played by a Lorentz-invariant evolution variable, which coincides with
  the transverse momentum of the latest emission in a suitably defined dipole
  zero-momentum frame. We also relate the evolution algorithm, which was used
  originally in the derivation of super-leading logarithms, to renormalization
  group evolution equations that have been encountered recently. Handling
  large colour matrices presents the most significant challenge to numerical
  implementations and we present a means to expand systematically about the
  leading colour approximation.}

\begin{document}

\maketitle
\flushbottom

\section{Introduction}

Precise predictions for observables at particle colliders often cannot be
achieved without the resummation of logarithmically enhanced
contributions to all orders in perturbation theory. Leading
logarithms of soft or collinear origin are accounted for in general
purpose event generators such as Herwig \cite{Bahr:2008pv,Bellm:2015jjp},
Pythia \cite{Sjostrand:2014zea} and Sherpa \cite{Gleisberg:2008ta}.
Resummation can be based on the direct analysis of contributing
Feynman graphs in QCD, and this is the approach we have taken in the past,
though effective field theories have also been
recognized as powerful tools to organize resummed calculations through 
a renormalization group evolution \cite{Becher:2014oda}.

For a large class of observables, which are fully inclusive below some
resolution scale in all phase-space regions, no contributions
originate from unresolved parton emission due to a perfect
cancellation of real and virtual corrections. This eases the
resummation procedure, as typically only one or very few emissions
need to be taken into account.  Observables
of this kind are referred to as global observables and include, for
example, many of the event shape variables measured at LEP.

On the other hand, observables that are insensitive to emissions into
a certain patch of phase space are called non-global and they are
subject to contributions from an arbitrary number of emissions,
effectively probing QCD dynamics in the blinded phase-space
region~\cite{Dasgupta:2001sh}. This effect makes the all-orders 
resummation (even of the leading contributions) a much more
complicated endeavour, mainly because non-trivial colour correlations
become unavoidable. Fortunately, these colour correlations simplify
dramatically in the leading colour approximation and they can
therefore be approximately accounted for in the general purpose event
generators.

In recent years there has been a good deal of progress in developing
the technology to tackle non-global observables and, in many cases, go
beyond the leading colour approximation
\cite{Weigert:2003mm,Khelifa-Kerfa:2015mma,Delenda:2015tbo,
   Schwartz:2014wha,
  Caron-Huot:2015bja,Larkoski:2015zka,Larkoski:2016zzc,Becher:2016mmh,
  Hatta:2013iba,Hagiwara:2015bia,Hatta:2017fwr}. Subleading
colour contributions have also been addressed in the context of parton
shower algorithms
\cite{Platzer:2012np,Nagy:2014mqa,Nagy:2015hwa,Nagy:2017ggp}. Often, attention
has focussed on processes with no coloured particles in the initial
state, not least because the simpler colour structure eliminates the
need to consider Coulomb (a.k.a. Glauber) gluons.  Coulomb gluon
interactions are particularly interesting since they have been shown
to induce a breakdown in the factorization of wide-angle soft gluon
emission from hard-collinear emission
\cite{Forshaw:2006fk,Forshaw:2008cq,Catani:2011st,Forshaw:2012bi,Schwartz:2017nmr}.

In spite of the existing progress, it remains to develop an automated
approach to resummation beyond the leading colour approximation for
general hard processes. Progress in this direction has been made by
Nagy and Soper \cite{Nagy:2014mqa,Nagy:2015hwa,Nagy:2017ggp}.
In this paper, our aim is to present a general framework that can be
used as a basis for future automated resummations. To be more
precise, we consider algorithmic, recursive definitions of QCD
amplitudes for the radiation of many soft gluons and
including leading virtual corrections to all orders.  Such an approach
is at the heart of direct QCD analyses of observables
involving many coloured legs, and it was used to identify the aforementioned
violations of strict collinear factorisation that occur at hadron
colliders.

The present work consists of two main parts, and a number of appendices
devoted to more technical details. In Section~\ref{sec:general} of the paper
we lay down the general evolution algorithm, in a form that is suited to the
calculation of multiple soft-gluon contributions to any observable in a
fully-differential way. We show how this approach connects to earlier work. We
then proceed to reformulate the algorithm in such a way as to make the
cancellation of infrared divergences explicit. We also highlight the role of
the specific ordering variable first identified in
\cite{Angeles-Martinez:2016dph}.

In Section~\ref{sec:colour}, we focus on the colour structures
encountered when solving the evolution equations. We present a
systematic procedure to calculate the resulting colour traces, which is 
based on the colour flow basis and the work presented
in \cite{Platzer:2013fha}. Identifying the leading
contribution leads us to re-derive the Banfi-Marchesini-Smye
equation \cite{Banfi:2010xy}. However our formalism is more general
and can systematically perform resummation of contributions enhanced
by the t'Hooft coupling $\alpha_s N\sim 1$, along with successive
perturbations that are parametrically suppressed by powers of $1/N$.

In Appendix~\ref{ap:others} we show how our approach connects to the
work presented by Becher et~al~\cite{Becher:2016mmh}, as well as
Weigert and Caron-Huot
\cite{Weigert:2003mm,Caron-Huot:2015bja,Caron-Huot:2016tzz}. In
Appendix~\ref{ap:BN} we make the cancellation of infrared divergences
explicit for observables that are inclusive below a resolution scale,
and in Appendix~\ref{app:fo} we explicitly calculate contributions in
a fixed-order expansion. Finally, Appendix~\ref{sec:dual} sets up
the machinery to deal with the fact that most colour bases are not
orthogonal (see \cite{Keppeler:2012ih} for a notable exception).

\section{The general algorithm}
\label{sec:general}

Our starting point is the cross section for emitting $n$ soft gluons. At this
stage we will assume that it is ok to order successive emissions in energy.
This assumption is ok for processes that are insensitive to Coulomb gluon
exchanges but appears not to be valid otherwise
\cite{Banfi:2010xy,Angeles-Martinez:2015rna}. We
have that

\bea
\sigma_0 &=&  \mathrm{Tr} ({\mathbf{V}}_{\mu,Q}\mathbf{H}(Q)
{\mathbf{V}}^\dag_{\mu,Q}) \equiv \mathrm{Tr} \; \mathbf{A}_0(\mu)
\nonumber \\
\dd\sigma_1 &=&  \mathrm{Tr}(  {\mathbf{V}}_{\mu,E_{1}}
\mathbf{D}_1^\mu
{\mathbf{V}}_{E_{1},Q}\mathbf{H}(Q){\mathbf{V}}^\dag_{E_{1},Q}\mathbf{D}_{1\mu}
^\dag {\mathbf{V}}^\dag_{\mu,E_{1}}) ~ \dd \Pi_1 \nonumber \\ &\equiv &
\mathrm{Tr} \; \mathbf{A}_1(\mu) \, \dd \Pi_1
\nonumber \\
\dd\sigma_2 &=&  \mathrm{Tr}(
 {\mathbf{V}}_{\mu,E_{2}} \mathbf{D}_2^\nu {\mathbf{V}}_{E_{2},E_{1}}
\mathbf{D}_1^\mu  {\mathbf{V}}_{E_{1},Q} \mathbf{H}(Q)
{\mathbf{V}}^\dag_{E_{1},Q}\mathbf{D}_{1\mu}^\dag
{\mathbf{V}}^\dag_{E_{2},E_{1}}\mathbf{D}^\dag_{2 \nu}
{\mathbf{V}}^\dag_{\mu,E_{2}}  ) ~ \dd \Pi_1 \dd \Pi_2 \nonumber \\ &\equiv &
\mathrm{Tr} \; \mathbf{A}_2(\mu) \, \dd \Pi_1 \dd \Pi_2
\hspace*{-1em} \nonumber \\ & \mathrm{ etc.} & \label{eq:sigma0} \eea
where $\mathbf{H}(Q)$ is the hard scattering matrix, $\mathbf{H} =
|\mathcal{M}\rangle \langle \mathcal{M}|$ and, in the eikonal
approximation,
\bea \mathbf{D}_i^\mu &=& \sum_j \bold{T}_j \; E_i
\frac{p_j^\mu}{p_j \cdot q_i} ~, \nonumber \\ \dd \Pi_i &=&
-\frac{\alpha_s}{\pi} \frac{\mu^{2\ep}\dd E_{i}}{E_{i}^{{1+2\ep }}}
\frac{\dd \Omega_{i}^{3-2\epsilon}}{4\pi {(2\pi)^{-2\ep}}
}~, \label{eq:simple}
\\ {\mathbf{V}}_{a,b} &=& \text{P} \exp \left[
  -\frac{\alpha_s {\mu^{2\ep}} }{\pi {(2\pi)^{-2\ep}}} \int_{a}^{b}
  \frac{\dd E_k}{E_k^{{1+2\ep}}} \sum_{i<j} (-\bold{T}_i \cdot
  \bold{T}_j) \, \left\{ \int \frac{\dd \Omega_k^{3-2\epsilon}}{4\pi }
  \;\omega_{ij}(\hat{k}) -i\pi \;
          {{\frac{\Omega^{2-2\epsilon}}{2\pi}}}\widetilde{\delta}_{ij}\right\}
          \right] ~, \nonumber \eea and\footnote{Note: $\Omega^d = 2
  \pi^{d/2}/\Gamma(d/2)$.}  \bea \omega_{ij}(\hat{k}) &=& E_k^2
\frac{(p_i \cdot p_j)}{(p_i \cdot k) \; (p_j \cdot k)}~.  \eea
If partons $i$ and $j$ are both in the initial state or they are both in
the final state then $\widetilde{\delta}_{ij}=1$ otherwise
$\widetilde{\delta}_{ij}=0$.  Note that the sum over partons $j$ in the
definition $\mathbf{D}_i^\mu$ is context-specific, i.e.\ it runs over
any prior soft gluon emissions in addition to the partons in the hard
scattering. Likewise, the colour charge operators, $\mathbf{T}_j$, and
the Sudakov operators, $\mathbf{V}_{a,b}$, are in a context-specific
representation of SU(3)$_c$. The operators $\mathbf{A}_n$ satisfy the
recurrence equation
\begin{eqnarray}
\mathbf{A}_n(E) = \mathbf{V}_{E,E_n}\mathbf{D}_n^\mu \; \mathbf{A}_{n-1}(E_n)
\; \mathbf{D}_{n\mu}^\dag \mathbf{V}_{E,E_n}^\dag \, \Theta(E \le E_n) ~, \label{eq:recur}
\end{eqnarray}
where $\Theta(E \le E_n)$ is the Heaviside function.
A general observable, $\Sigma$, can be computed using
\begin{eqnarray}
	\Sigma(\mu) &=& \int \sum_n ~ \dd \sigma_n \,
u_n(k_1,k_2,\cdots,k_n)~, \label{eq:eord} \\
&=& \sum_n  \left(\prod_{i=1}^n  \int \dd \Pi_i  \right)\,\text{Tr}\,  \mathbf{A}_n(\mu) ~
u_n(k_1,k_2,\cdots k_n) ~, \nonumber
\end{eqnarray}
where the $u_n$ are the observable dependent measurement functions and
the $k_i$ are soft gluon momenta. We suppress dependence on the hard
partons and integration over their phase space.

In the above, we should take the limit $\mu \to 0$, though we will
consider non-zero values in what follows.  The carat on $\hat{k}$
reminds us that $\omega_{ij}(\hat{k})$ is dependent only upon the
direction of the vector $k$ in the $ij$ rest frame. The path-ordering,
P, in the definition of $\mathbf{V}_{a,b}$ is not actually needed
here, because the expression in curly brackets in
Eq.~(\ref{eq:simple}) is independent of the ordering variable, $E_k$.
The cross sections in Eq.~(\ref{eq:sigma0}) are the general building
blocks for any observable and they can be used as the basis for a
Monte Carlo computer code to generate partonic events.

We can also write an evolution equation\footnote{For simplicity, we work in
$d=4$ spacetime dimensions unless otherwise stated.}:
\bea
E\frac{\partial \mathbf{A}_n(E)}{\partial E} &=& -\boldsymbol{\Gamma}
\mathbf{A}_n(E) - \mathbf{A}_n(E)\boldsymbol{\Gamma}^\dag +  \mathbf{D}_n^\mu
\;  \mathbf{A}_{n-1}(E) \; \mathbf{D}_{n\mu}^\dag \; E\,\delta(E - E_n)
~,\label{eq:evo1}
\eea
where
\beq
\boldsymbol{\Gamma} = \frac{\alpha_s}{\pi}\sum_{i<j} (-\bold{T}_i \cdot
\bold{T}_j)
\;  \left\{ \int \frac{\dd \Omega_k}{4\pi} \;\omega_{ij}(\hat{k})  -i\pi \;
\widetilde{\delta}_{ij}\right\} ~. \label{eq:Gamma}
\eeq
When the measurement function factorizes, i.e.
$u_n(k_1,k_2,\cdots k_n) = u(E_1,\hat{k}_1) \cdots u(E_n,\hat{k}_n)$, we
can define
\bea
\mathbf{G}_n(E) = \prod_{i=1}^n \int \frac{\dd E_i}{E_i} \, \mathbf{A}_n(E) ~
\eea
and Eq.~({\ref{eq:evo1}) becomes
\bea
E\frac{\partial \mathbf{G}_n(E)}{\partial E} &=& -\boldsymbol{\Gamma}
\mathbf{G}_n(E) - \mathbf{G}_n(E)\boldsymbol{\Gamma}^\dag +  \mathbf{D}_n^\mu
\;  \mathbf{G}_{n-1}(E) \; \mathbf{D}_{n\mu}^\dag~u(E,\hat{k}_n).  \label{eq:evo2}
\eea
This can be re-written as
\bea
E\frac{\partial \mathbf{G}_n(E)}{\partial E} &=&
\frac{\alpha_s}{\pi} \sum_{i<j} \Bigg[  \int \frac{\dd \Omega_k}{4 \pi}
\omega_{ij}(\hat{k}) \Big( \mathbf{T}_i \cdot \mathbf{T}_j \, \mathbf{G}_n
+  \mathbf{G}_n \, \mathbf{T}_i \cdot \mathbf{T}_j \Big)  \label{eq:evo3}\\ & & - ~
\omega_{ij}(\hat{k}_n)(\mathbf{T}_i \, \mathbf{G}_{n-1} \mathbf{T}_j^\dag + \mathbf{T}_j \,
\mathbf{G}_{n-1} \mathbf{T}_i^\dag) \, u(E,\hat{k}_n) \Bigg] ~~ + \text{~~Coulomb
terms.} \nonumber
\eea
In Appendix \ref{ap:others}, we show that
Eq.~(\ref{eq:evo3}) is the same as the leading-logarithmic accuracy RG equations considered in
\cite{Weigert:2003mm,Becher:2016mmh,Caron-Huot:2015bja,Caron-Huot:2016tzz}.

Note that, if we are interested in a specific observable that we know is fully
inclusive of real emissions with $E < Q_0$ then we can use the Bloch-Nordsieck
cancellation in order to fix $\mu = Q_0$ and integrate the real-emission phase
space over $E > Q_0$. This is proved in Appendix
\ref{ap:BN}.

\subsection{An infra-red finite reformulation \label{sec:ngo}}
As it stands, there are fixed-order, infra-red divergences in $\mathbf{A}_n$ that
cancel in the sum over $n$ after integration over the 
real emissions. In this section we present a reformulation of
the algorithm in which the cancellation of infra-red divergences (both
soft and soft-collinear) arising in the eikonal approximation is
manifest.

General observables may be defined by dividing the
angular phase-space into two complementary sub-regions, which we refer to as
the "in" and "out" regions, such that the observable is fully
inclusive over emissions in the "out" region. If the "out" region is of zero
extent then the observable is referred to as a global observable, otherwise it
is known as a non-global observable. Phrased this way, we see that all
observables are non-global to some extent (since $4\pi$-detectors do not
exist).

In order to expose the infra-red cancellation, it is useful to break apart the
virtual loop-integral in $\mathbf{\Gamma}$ (see Eq.~(\ref{eq:Gamma})) so that
we expose the part which is destined to cancel against a corresponding real
emission contribution. To this end, it is useful to consider the measurement
function in the soft gluon limit:
\beq
u_m(q_1,\cdots,q_m) = u(q_j,\{ q_1,\cdots,q_{j-1},q_{j+1},\cdots,
q_m\}) \, u_{m-1}(q_1,\cdots,q_{j-1},q_{j+1},\cdots,
q_m)~.
\eeq
This is quite general, the important thing is that, $u(q_j,\{
q_1,\cdots,q_{j-1},q_{j+1},\cdots,
q_m\}) \to 1$ in the limit that gluon $j$ has zero energy.
	Generally, we can
write
\begin{equation}
	u(k,\{q\}) = \Theta_{\text{out}}(k) + \Theta_{\text{in}}(k)
	u_{\text{in}}(k,\{ q \}) ~.
\end{equation}
The set $\{q\}$ corresponds to all other real emissions and 
$\Theta_{\text{in/out}}(k)$ is defined to be unity if $k$ is in the "in"/"out" 
region and zero otherwise. For global observables, the "out" region is 
of zero extent, in which case $u(k,\{q\}) = u_\text{in}(k,\{q\})$. 
Armed with this we define
\begin{align}
\begin{gathered}
\bd{\Gamma} = \bd{\Gamma}_{u}+\wb{\bd{\Gamma}}_{u},\\
\wb{\bd{\Gamma}}_{u}=  \frac{\alpha_s}{\pi}   \left( \int \frac{\dd
\Omega_k}{4\pi}
\, (1-u(k,\{q\})) \frac{\mathbf{D}^2_k}{2} + i\pi\sum_{i< j}
\wt{\delta}_{ij}\bold{T}_i \cdot \bold{T}_j \right) ~, \\
\bd\Gamma_{u}=  \frac{\alpha_s}{\pi}   \int \frac{\dd \Omega_k}{4\pi} \,
u(k,\{q\})
\frac{\mathbf{D}^2_k}{2}, \\
\mathbf{V}_{a,b} = \wb{\mathbf{V}}_{a,b} - \int \dd \Pi_1 \, u(k_1,\{q\})
\,\wb{\mathbf{V}}_{a,1} \,
\frac{\mathbf{D}_1^2}{2} \, \wb{\mathbf{V}}_{1,b} \\
+(-1)^2  \int \dd \Pi_2 \, \dd \Pi_1\,\Theta(E_2<E_1) \,  u(k_1,\{q\}) \,
u(k_2,\{q\}) \,
\wb{\mathbf{V}}_{a,2} \,
\frac{\mathbf{D}_2^2}{2} \, \wb{\mathbf{V}}_{2,1} \,
\frac{\mathbf{D}_1^2}{2}  \,  \wb{\mathbf{V}}_{1,b} + \cdots ,\\
\wb{\mathbf{V}}_{a,b}\equiv \text{P}\, \text{exp}\left\{\int_{a}^{b} \frac{\dd
E}{E} \, \wb{\bd\Gamma}_{u}  \right\} ~,
\end{gathered}
\end{align}
where \bea \frac{1}{2}\mathbf{D}_a^2 &=& \sum_{i<j} (-\bold{T}_i \cdot
\bold{T}_j) \; \omega_{ij}(\hat{k}_a) ~ .  \eea Notice that the
virtual gluons are summed to all orders only if they are in the "in"
region, i.e.\ $\wb{\mathbf{V}}$ involves virtual gluons integrated over
the "in" region. Since $1-u(k,\{q\}) \to 0$ when $E_k \to 0$ it
follows that there are no soft singularities in $\wb{\mathbf{V}}$
except those arising from Coulomb gluon exchange. The poles from
Coulomb gluon exchange cancel though because they always appear in
terms $\sim \mathrm{Tr}(\wb{\mathbf{V}}_{0,a} \cdots
\wb{\mathbf{V}}_{0,a}^\dag)$. The cyclicity of the trace ensures that
the $i \pi$ terms can be combined into the unit matrix, since the real
part of $\wb{\mathbf{V}}_{0,a}$ vanishes in the limit $E \to 0$.

We can now re-write the
observable as
\beq
 \Sigma(\mu) =\sum_n  \left(\prod_{i=1}^n  \int \dd \Pi_i  \right)\,
   \text{Tr}\, \bd{B}_{n}(\mu)~ \Phi_n(q_1,q_2,\cdots,q_n)~,
 \label{eq:nonglobalexpansion}
\eeq
where the operators $\bd{B}_{n}$
satisfy the recurrence relation (i.e.\ the analogue of
Eq.~(\ref{eq:recur})):
\begin{eqnarray}
	\bd{B}_{n} (E)&=& \wb{\bd{V}}_{E,E_n}\Bigg[
\bd{D}_n^\mu \bd{B}_{n-1} (E_n)\bd{D}_{n \mu}^\dag \delta_n^{R} \\ & &
-\left\{\bd{B}_{n-1} (E_n), \frac{\bd{D}_n^2}{2} \right\} \delta_{n}^{V}
u(q_n,\{ q \})
\Bigg] \,
\wb{\bd{V}}_{E,E_n}^\dagger \, \Theta(E \le E_n) ~. \nonumber \label{eq:Brec}
\end{eqnarray}
It should be understood that $\{\delta_{i}^{R}=1, \delta_{i}^{V}=0\}$ if $i$
is a real emission and $\{\delta_{i}^R=0,\delta_{i}^V=1\}$ if it is
virtual. Also, $\bd{B}_0(E)= \wb{\bd{V}}_{E,Q}\,\bd{H}\,
\wb{\bd{V}}^\dagger_{E,Q} $, and $\mu$ should be set equal to zero. In
Eq.~(\ref{eq:nonglobalexpansion}), $\Phi_n$ encodes the measurement functions
for any number of real emissions:
\begin{align}
&\Phi_0\equiv 1, \nonumber\\
&\Phi_n (q_1,q_2,\dots,q_n)\equiv
 \sum_{a=1}^{2^n} u_m(\{q_k\})\Bigg|_{\substack{k \in P_n^a,\\ m = |P_n^a|}}
\left( \prod_{j \in P_n^a}\delta_{j}^{R}\right)
\left( \prod_{i \notin P_n^a}\delta_{i}^{V}\right) .
\end{align}
The $P_n^a$ set is indexed by $a$ indicating which of the $n$ gluons are real,
and $m$ is the cardinality of the set. For example, if $n=2$ then
$P_2^1 = \{1,2\}$ indicates that gluons $q_1$ and $q_2$ are both
real and $m=2$, $P_2^2 = \{1\}$ indicates that $q_1$ is real and $q_2$
is
virtual ($m=1$),
$P_2^3
= \{2\}$ indicates that $q_2$ is real and $q_1$ is virtual ($m=1$), and
$P_2^4 = \{ \}$ indicates that both gluons are virtual ($m=0$). In this case,
\begin{align}
\Phi_2 (q_1,q_2)= \delta_{1}^{V} \delta_{2}^{V}
+u_1(q_1) \delta_{1}^{R}\delta_{2}^{V} + u_1(q_2) \delta_{2}^{R}
\delta_{1}^{V}
+u_2(q_1,q_2)\delta_{1}^{R} \delta_{2}^{R}~
\end{align}
and, recalling that $u(q,\{ \}) = u_1(q)$ and $u(q,\{ q_1 \})
u_1(q_1) = u_2(q_1,q)$, \bea \Sigma_0 &=& \mathrm{Tr}
(\wb{\bd{V}}_{0,Q}\,\bd{H}\, \wb{\bd{V}}^\dagger_{0,Q} ) ~,
\\ \Sigma_1 &=& \int \dd \Pi_1 ~ u_1(q_1) \, \text{Tr} \Big(
\wb{\bd{V}}_{0,E_1} \mathbf{D}_1^\mu \wb{\bd{V}}_{E_1,Q} \mathbf{H}
\wb{\bd{V}}_{E_1,Q}^\dag \mathbf{D}_{1\mu}^\dag
\wb{\bd{V}}_{0,E_1}^\dag \nonumber \\ & & - \wb{\bd{V}}_{0,Q}
\mathbf{H} \wb{\bd{V}}_{E_1,Q}^\dag \frac{\mathbf{D}_1^2}{2}
\wb{\bd{V}}_{0,E_1}^\dag -
\wb{\bd{V}}_{0,E_1}\frac{\mathbf{D}_1^2}{2}\wb{\bd{V}}_{E_1,Q}
\mathbf{H} \wb{\bd{V}}_{0,Q}^\dag \Big) ~, \nonumber \\ \Sigma_2 &=&
\int \mathrm{d} \Pi_1 \mathrm{d} \Pi_2 ~ \Big( u_1(q_1) u_1(q_2)
\left( \Sigma_2^{VR} + \Sigma_2^{VV} \right) + u_2 (q_1,q_2)
(\Sigma_2^{RR} + \Sigma_2^{RV})\Big) ~ , \nonumber \eea where \bea
\Sigma_2^{RR} &=& \text{Tr} \left( \wb{\bd{V}}_{0,E_2}
\mathbf{D}_2^\nu \wb{\bd{V}}_{E_2,E_1} \mathbf{D}_1^\mu
\wb{\bd{V}}_{E_1,Q} \mathbf{H} \wb{\bd{V}}_{E_1,Q}^\dagger
\mathbf{D}_{1\mu}^\dagger \wb{\bd{V}}_{E_2,E_1}^\dagger
\mathbf{D}_{2\nu}^\dagger \wb{\bd{V}}_{0,E_2}^\dagger \right )
\\ \Sigma_2^{VR} &=& - \text{Tr} \left( \wb{\bd{V}}_{0,E_2}
\mathbf{D}_2^\mu \wb{\bd{V}}_{E_2,Q} \mathbf{H}
\wb{\bd{V}}_{E_1,Q}^\dagger \frac{\mathbf{D}_1^2}{2}
\wb{\bd{V}}_{E_2,E_1}^\dagger \mathbf{D}_{2 \mu}^\dagger
\wb{\bd{V}}_{0,E_2}^\dagger \right. \nonumber \\ &+&
\left. \wb{\bd{V}}_{0,E_2} \mathbf{D}_2^\mu \wb{\bd{V}}_{E_2,E_1}
\frac{\mathbf{D}_1^2}{2} \wb{\bd{V}}_{E_1,Q} \mathbf{H}
\wb{\bd{V}}_{E_2,Q}^\dagger \mathbf{D}_{2 \mu}^\dagger
\wb{\bd{V}}_{0,E_2}^\dagger \right) \nonumber \\ \Sigma_2^{RV} &=& -
\text{Tr} \left( \wb{\bd{V}}_{0,E_1} \mathbf{D}_1^\mu
\wb{\bd{V}}_{E_1,Q} \mathbf{H} \wb{\bd{V}}_{E_1,Q}^\dagger
\mathbf{D}_{1 \mu}^\dagger \wb{\bd{V}}_{E_2,E_1}^\dagger
\frac{\mathbf{D}_2^2}{2} \wb{\bd{V}}_{0,E_2}^\dagger \right. \nonumber
\\ & & + \left. \wb{\bd{V}}_{0,E_2} \frac{\mathbf{D}_2^2}{2}
\wb{\bd{V}}_{E_2,E_1} \mathbf{D}_1^\mu \wb{\bd{V}}_{E_1,Q} \mathbf{H}
\wb{\bd{V}}_{E_1,Q}^\dagger \mathbf{D}_{1 \mu}^\dagger
\wb{\bd{V}}_{0,E_1}^\dagger \right ) \nonumber \\ \Sigma_2^{VV} &=&
\left(\wb{\bd{V}}_{0,Q} \mathbf{H} \wb{\bd{V}}_{E_1,Q}^\dagger
\frac{\mathbf{D}_1^2}{2} \wb{\bd{V}}_{E_2,E_1}^\dagger +
\wb{\bd{V}}_{0,E_1} \frac{\mathbf{D}_1^2}{2} \wb{\bd{V}}_{E_1,Q}
\mathbf{H} \wb{\bd{V}}_{E_2,Q}^\dagger \right)
\frac{\mathbf{D}_2^2}{2} \wb{\bd{V}}_{0,E_2}^\dagger \nonumber \\ & &
+ \wb{\bd{V}}_{0,E_2} \frac{\mathbf{D}_2^2}{2} \left(
\wb{\bd{V}}_{E_2,Q} \mathbf{H} \wb{\bd{V}}_{E_1,Q}^\dagger
\frac{\mathbf{D}_1^2}{2} \wb{\bd{V}}_{0,E_1}^\dagger +
\wb{\bd{V}}_{E_2,E_1} \frac{\mathbf{D}_1^2}{2} \wb{\bd{V}}_{E_1,Q}
\mathbf{H} \wb{\bd{V}}_{0,Q}^\dagger \right)~. \nonumber \eea Note
that in $\Sigma_2$ the pairs of contributions in which the softer
gluon is real or virtual (i.e.\ $\Sigma_2^{VR} + \Sigma_2^{VV}$ and
$\Sigma_2^{RR} + \Sigma_2^{RV}$) are proportional to the same
expression, ensuring that their soft singularities cancel.  For an
infra-red safe observable the $\Sigma_n$ are all finite (in the
eikonal approximation). In Appendix \ref{app:fo} we show how the
cancellation of infra-red poles works out by explicit calculation of
the non-global contribution to the hemisphere mass in $e^+ e^-$
collisions to order $\alpha_s^3$.

\subsubsection{Non-global observables: a simple example \label{sec:examples}}

In the leading logarithm approximation, many observables can be
computed with a factorizable measurement function, i.e.
\begin{align}
u_n(q_1,\dots,q_n)& = \prod_{i=1}^n \, u_1(q_i)~.
\end{align}
Also, in many cases (such as the hemisphere jet mass and
gaps-between-jets), the measurement function simply vetoes real
emissions into some region of phase-space, e.g.  \beq u_1(q) =
\Theta_{\text{out}}(q) + \Theta_{\text{in}}(q)\, \Theta(\rho > E)~.  \eeq In
these cases, we can simply set $\mu = \rho$ in
Eq.~(\ref{eq:nonglobalexpansion}), take \beq
\wb{\mathbf{V}}_{a,b}\equiv \text{P}\, \text{exp}\left\{
\frac{\alpha_s}{\pi} \int_{a}^{b} \frac{\dd E}{E} \left(
\int_\text{in} \frac{\dd \Omega_k}{4\pi} \, \frac{\mathbf{D}^2_k}{2} +
i\pi\sum_{i< j} \wt{\delta}_{ij}\bold{T}_i \cdot \bold{T}_j \right)
\right\} \eeq and replace $u(q_n,\{q\}) \to \Theta_{\text{out}}(q_n)$
in Eq.~(\ref{eq:Brec}). This is because of the inclusivity of the
observable for $E < \rho$, which leads to a complete cancellation of
the real and virtual contributions (see Appendix \ref{ap:BN}). In
other words, $\Sigma_n$ is the contribution from $n$ gluons in the
"out" region (the gluons can be real or virtual).  Specifically,
\bea \label{eq:sngl} \Sigma_0 &=& \mathrm{Tr}
(\wb{{\mathbf{V}}}_{\rho,Q}\mathbf{H}\wb{{\mathbf{V}}}^{\dag}_{\rho,Q})
\\ \Sigma_1 &=& \int_{\text{out}} \dd \Pi_1 \, \mathrm{Tr} \Big(
\wb{{\mathbf{V}}}_{\rho,E_{1}} \mathbf{D}_1^\mu
\wb{\mathbf{V}}_{E_{1},Q}\mathbf{H}\wb{{\mathbf{V}}}^{\dag}_{E_{1},Q}
\mathbf{D}_{1\mu}^\dag \wb{{\mathbf{V}}}^{\dag}_{\rho,E_{1}} ~
\nonumber \\ & & - \wb{\bd{V}}_{\rho,Q} \mathbf{H}
\wb{\bd{V}}_{E_1,Q}^\dag \frac{\mathbf{D}_1^2}{2}
\wb{\bd{V}}_{\rho,E_1}^\dag - \wb{{\mathbf{V}}}_{\rho,E_{1}}
\frac{\mathbf{D}_1^2}{2}
\wb{{\mathbf{V}}}_{E_{1},Q}\mathbf{H}\wb{{\mathbf{V}}}^{\dag}_{\rho,Q}
\Big) \nonumber \\ \Sigma_2 &=& \int_{\text{out}} \dd \Pi_1 \,
\int_{\text{out}} \dd \Pi_2 \, \mathrm{Tr} \left(
\wb{\bd{V}}_{\rho,E_2} \mathbf{D}_2^\nu \wb{\bd{V}}_{E_2,E_1}
\mathbf{D}_1^\mu \wb{\bd{V}}_{E_1,Q} \mathbf{H}
\wb{\bd{V}}_{E_1,Q}^\dagger \mathbf{D}_{1\mu}^\dagger
\wb{\bd{V}}_{E_2,E_1}^\dagger \mathbf{D}_{2\nu}^\dagger
\wb{\bd{V}}_{\rho,E_2}^\dagger \right.  \nonumber \\ && -
\left. \wb{\bd{V}}_{\rho,E_2} \mathbf{D}_2^\mu \wb{\bd{V}}_{E_2,Q}
\mathbf{H} \wb{\bd{V}}_{E_1,Q}^\dagger \frac{\mathbf{D}_1^2}{2}
\wb{\bd{V}}_{E_2,E_1}^\dagger \mathbf{D}_{2 \mu}^\dagger
\wb{\bd{V}}_{\rho,E_2}^\dagger \right. \nonumber \\ && -
\left. \wb{\bd{V}}_{\rho,E_2} \mathbf{D}_2^\mu \wb{\bd{V}}_{E_2,E_1}
\frac{\mathbf{D}_1^2}{2} \wb{\bd{V}}_{E_1,Q} \mathbf{H}
\wb{\bd{V}}_{E_2,Q}^\dagger \mathbf{D}_{2 \mu}^\dagger
\wb{\bd{V}}_{\rho,E_2}^\dagger \right. \nonumber \\ && -
\left. \wb{\bd{V}}_{\rho,E_1} \mathbf{D}_1^\mu \wb{\bd{V}}_{E_1,Q}
\mathbf{H} \wb{\bd{V}}_{E_1,Q}^\dagger \mathbf{D}_{1 \mu}^\dagger
\wb{\bd{V}}_{E_2,E_1}^\dagger \frac{\mathbf{D}_2^2}{2}
\wb{\bd{V}}_{\rho,E_2}^\dagger \right. \nonumber \\ &&
-\left. \wb{\bd{V}}_{\rho,E_2} \frac{\mathbf{D}_2^2}{2}
\wb{\bd{V}}_{E_2,E_1} \mathbf{D}_1^\mu \wb{\bd{V}}_{E_1,Q} \mathbf{H}
\wb{\bd{V}}_{E_1,Q}^\dagger \mathbf{D}_{1 \mu}^\dagger
\wb{\bd{V}}_{\rho,E_1}^\dagger \right . \nonumber \\ && +
\left. \wb{\bd{V}}_{\rho,Q} \mathbf{H} \wb{\bd{V}}_{E_1,Q}^\dagger
\frac{\mathbf{D}_1^2}{2} \wb{\bd{V}}_{E_2,E_1}^\dagger
\frac{\mathbf{D}_2^2}{2} \wb{\bd{V}}_{\rho,E_2}^\dagger +
\wb{\bd{V}}_{\rho,E_1} \frac{\mathbf{D}_1^2}{2} \wb{\bd{V}}_{E_1,Q}
\mathbf{H} \wb{\bd{V}}_{E_2,Q}^\dagger \frac{\mathbf{D}_2^2}{2}
\wb{\bd{V}}_{\rho,E_2}^\dagger \right. \nonumber \\ && +
\left. \wb{\bd{V}}_{\rho,E_2} \frac{\mathbf{D}_2^2}{2}
\wb{\bd{V}}_{E_2,Q} \mathbf{H} \wb{\bd{V}}_{E_1,Q}^\dagger
\frac{\mathbf{D}_1^2}{2} \wb{\bd{V}}_{\rho,E_1}^\dagger +
\wb{\bd{V}}_{\rho,E_2} \frac{\mathbf{D}_2^2}{2} \wb{\bd{V}}_{E_2,E_1}
\frac{\mathbf{D}_1^2}{2} \wb{\bd{V}}_{E_1,Q} \mathbf{H}
\wb{\bd{V}}_{\rho,Q}^\dagger \right)~. \nonumber \\ & \mathrm{ etc.} &
\nonumber \label{eq:sigma} \eea This is the "out of gap" expansion
used in \cite{Forshaw:2006fk,Forshaw:2008cq} to derive the
super-leading logarithmic contribution to gaps-between-jets.

\subsection{The ordering variable} \label{sec:ordering}
So far we have presumed energy ordering in the virtual gluon operators
$\mathbf{V}_{a,b}$. However, this is known not to generate the correct
super-leading logarithms and instead transverse momentum ordering
should be used \cite{Angeles-Martinez:2015rna}.
Interestingly, and working in the eikonal
approximation, but {\it only} for gluons that couple to the original
hard partons, explicit calculation of all relevant Feynman diagrams
(at one loop) reveals that the corrections associated with the exact
triple and four-gluon vertices can be largely subsumed into a Lorentz
invariant ordering variable, in a potentially simple extension of the
algorithm that we described
in~\cite{Angeles-Martinez:2015rna,Angeles-Martinez:2016dph}. This
intriguing physics may not be so clearly visible in an effective field
theory treatment, where the evolution is in an arbitrary
renormalization scale.

\begin{figure}
    \centering
    \begin{subfigure}[c]{0.3\textwidth}
        \includegraphics[width=\textwidth]{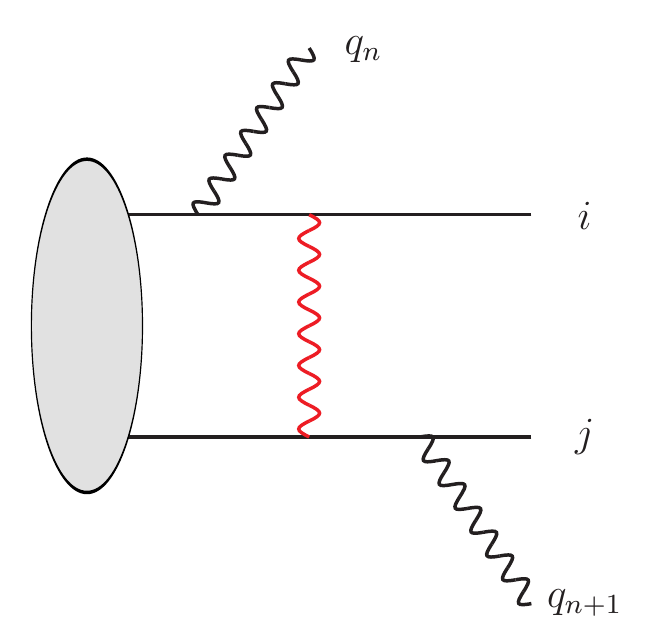}
    \end{subfigure}
    ~
    \begin{subfigure}[c]{0.3\textwidth}
      $$ \int_{q_{n+1}^{(ij)}}^{{q_{n}^{(ij)}}} \frac{\mathrm{d} k_T}{k_T}$$
    \end{subfigure}

    \begin{subfigure}[c]{0.3\textwidth}
        \includegraphics[width=\textwidth]{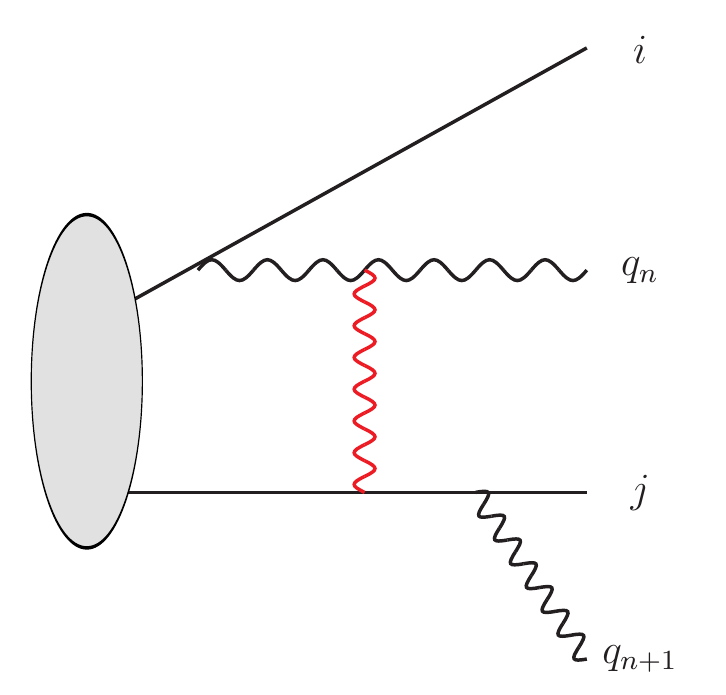}
    \end{subfigure}
    ~
    \begin{subfigure}[c]{0.3\textwidth}
        $$ \int_{q_{n+1}^{([n]j)}}^{{q_{n}^{(ij)}}} \frac{\mathrm{d} k_T}{k_T}$$
    \end{subfigure}
    \caption{\label{fig:dipolept} Illustrating how the `dipole transverse
      momentum' serves to limit the virtual loop integration. The index $[n]$
      refers to the gluon with momentum $q_n$. }
\end{figure}

Figure~\ref{fig:dipolept} illustrates the key finding of
\cite{Angeles-Martinez:2016dph}. The upper graph illustrates that the virtual
gluon loop integral has its transverse momentum limited by the transverse
momenta of the two nearest real emissions.  The relevant transverse momentum
of a gluon, $q^{(ij)}$, is defined by its Sudakov decomposition over the
momenta $p_i$ and $p_j$ involved in dipole $(ij)$, and is given by \beq
(q^{(ij)})^2 = \frac{2 q\cdot p_i \; q \cdot p_j}{p_i \cdot p_j} = \frac{2
  E^2}{\omega_{ij}(\hat{q})}~. \label{eq:dipdef1} \eeq When one of the gluons
to which the virtual gluon couples also happens to be one of the nearest
emissions, the relevant dipole transverse momentum vanishes. In this case
however, the explicit calculation reveals that the relevant dipole momentum is
that of the {\em parent} of the parton which couples to the virtual gluon,
i.e.\ parton $i$ in the lower graph of Figure~\ref{fig:dipolept}.

The corresponding differential cross section
has a similar structure to Eq.~\eqref{eq:sigma0} but with
\bea
\bold{A}_0 (\tilde{q}_1) &\equiv &  \bold{H}(Q)+
\sum_{i<j}
\left(
\bold{I}_{ij} \left(\widetilde{q}_{1} , Q \right)
\bold{H}(Q)+
\bold{H}(Q) \, \bold{I}_{ij}^\dagger \left(\widetilde{q}_{1} , Q \right)
\right) ,
\nonumber \\
\bold{A}_n(\widetilde{q}_{n+1}) &\equiv &
 \sum_{k\ne l} \Bigg( \bold{T}_{k}\, \bold{A}_{n-1}(\widetilde q_n)\,
 \bold{T}_{l}\,\omega_{kl}(\hat{q}_n) +
 \Big[ \sum_{i<j \ne [n]}
 \bold{I}_{ij} \left(\widetilde{q}_{n+1} , \widetilde{q}_{n} \right)
 \bold{T}_{k}\,
 \bold{A}_{n-1}(\widetilde q_n)\, \bold{T}_{l}\, \omega_{kl}(\hat{q}_n)
 \label{eq:dkto} \nonumber \\
 & + & \sum_{i\ne [n]} \bold{I}_{i[n]}\!\! \left(\widetilde{q}_{n+1} , q_n^{(i
   k)}\right) \bold{T}_{k}\, \bold{A}_{n-1}(\widetilde q_n)\, \bold{T}_{l}\,
 \left(\omega_{kl}(\hat{q}_n)- \omega_{il}(\hat{q}_n) \right) +\text{h.c.}
 \Big] \Bigg)~, \eea where it should be understood that
 $\widetilde{q}=q^{(ij)}$ when placed in the argument of $\bold{I}_{ij}$. The
 Lorentz invariant operator $\bold{I}_{ij}(a,b)$ is given by \beq
 \mathbf{I}_{ij}(a,b) = \frac{\alpha_s }{\pi} \bold{T}_i \cdot \bold{T}_j
 \,\int_{a}^{b} \frac{\dd k_T}{k_T} \left\{ \int \frac{\dd \Omega}{4\pi } \,
 \omega_{ij}(\hat{k}) \, \theta_{ij}(k) -i\pi \;
 \widetilde{\delta}_{ij}\right\}, \eeq where $\theta_{ij}(k) = \Theta(p_j
 (p_i-k) > 0) \Theta(p_i (p_j-k) > 0)$ restricts the region of the angular
 integration to be the same as in the phase space integral for a real gluon
 with the same transverse momentum.  This can be written \beq
 \left|\ln\tan\frac{\theta}2\right| < \ln\frac{\sqrt{2p_i\ldot p_j}}{k_T}\,,
 \qquad\mbox{or}\quad |\sin\theta|\gtap\frac{2k_T}{\sqrt{2p_i\ldot p_j}}~.
 \eeq Up to the limits on the $\cos \theta$ integral, $\mathbf{I}_{ij}$ is
 equal to the exponent in the $\mathbf{V}$ operator defined in
 Eq.~\eqref{eq:simple}. In \cite{Angeles-Martinez:2016dph} we presented the
 form of this operator in $4- 2\epsilon$ dimensions after integration over the
 solid angle, i.e.  \bea
\label{eq:insertop}
	\mathbf{I}_{ij}(a,b) &=& \frac{\alpha_s}{2 \pi}
        \frac{c_{\Gamma}}{\epsilon^2} \, \mathbf{T}_i \cdot
        \mathbf{T}_j \Bigg[ \left( \frac{b^2}{4 \pi \mu^2}
          \right)^{-\epsilon} \left(1 +i \pi \epsilon \,
          \tilde\delta_{ij}-\epsilon\ln\frac{2p_i\cdot
            p_j}{b^2}\right) \nonumber \\ & & - \left(\frac{a^2}{4\pi
            \mu^2} \right)^{-\epsilon} \, \left(1 +i \pi \epsilon \,
          \tilde\delta_{ij}-\epsilon\ln\frac{2p_i\cdot
            p_j}{a^2}\right) \Bigg]~, \eea where $c_{\Gamma} = 1 -
        \epsilon \gamma_E$, and $\gamma_E$ is the Euler-Mascheroni
        constant. This expression is accurate up to non-logarithmic
        terms of order $\epsilon^0$ in the real part and order
        $\epsilon^1$ in the imaginary part. When both scales $a$ and
        $b$ are non-zero, $\mathbf{I}_{ij}$ is finite and given by
\begin{eqnarray}
	\mathbf{I}_{ij}(a,b) &=&  \frac{\alpha_s}{2 \pi}
 \, \mathbf{T}_i \cdot \mathbf{T}_j
\biggl[ -\frac12\ln^2\frac{2p_i\cdot p_j}{b^2}
 +\frac12\ln^2\frac{2p_i\cdot p_j}{a^2}
 -i \pi \tilde\delta_{ij}\ln\frac{b^2}{a^2} \biggr]~.
\end{eqnarray}
Eq.~(\ref{eq:insertop}) would be identical to the result of Catani and
Grazzini \cite{Catani:2000pi} (see also
\cite{Bern:1999ry,Feige:2014wja}) if we replaced the factor $1 +i \pi
\epsilon \,\tilde\delta_{ij}$ in Eq.~(\ref{eq:insertop}) by $\cos(\pi
\epsilon) + i \sin(\pi \epsilon \tilde\delta_{ij})$.  It should be
stressed that the calculations in
\cite{Angeles-Martinez:2015rna,Angeles-Martinez:2016dph} were
performed only at one loop and it remains to be seen how the improved
resummation proceeds to all orders.

\section{Large-$\boldsymbol{N}$ structures}
\label{sec:colour}

In practical calculations, the colour algebra rapidly becomes intractable
after only a few real gluon emissions. To simplify matters, we shall now
identify the leading contributions in an expansion in the number of
colours. We will work in the colour flow basis \cite{Maltoni:2002mq}, which is
closely related to the way colour is treated in parton shower algorithms
\cite{Gustafson:1987rq,Lonnblad:1992tz,Platzer:2009jq,Hoche:2015sya}.  In the
following subsection, we will show that, to leading order in the number of
colours, our algorithm gives rise to a dipole-type parton shower and that it
reproduces the Banfi-Marchesini-Smye equation \cite{Banfi:2002hw}. We then
turn our attention to setting up a framework to calculate the first
subleading-colour corrections.

\subsection{Colour flow basis}

To start with, we collect together some of the key results concerning
the colour flow basis.  We label the set of basis tensors as
$\{|\sigma\rangle\}$, and we assign a colour or anti-colour index,
$c_i$ or $\bar{c}_i$, to each external leg $i$ of any scattering
amplitude. Gluons carry both colour and anti-colour and incoming
quarks carry anti-colour. We start to count colour index labels from
$1$, and choose $c_i=0$ ($\bar{c}_i=0$) if $i$ only carries
anti-colour (only carries colour). The basis tensors are labelled by
permutations $\sigma$ of the colour indices and are given by products
of Kronecker $\delta$'s as
\begin{equation}
|\sigma\rangle =\left|
\begin{array}{ccc}
1 & \cdots & n \\
\sigma(1) &\cdots & \sigma(n)
\end{array}
\right\rangle  = \delta^{\alpha_1}_{\bar{\alpha}_{\sigma(1)}}\cdots
\delta^{\alpha_n}_{\bar{\alpha}_{\sigma(n)}} \ ,
\end{equation}
where the $\alpha_{1...n}$ and $\bar{\alpha}_{1...n}$ are fundamental
and anti-fundamental indices assigned to the colour (anti-colour)
legs, taking values in the actual number of colours $1,...,N$. There
are $n= n_q + n_g = n_{\bar{q}} + n_g$ possible colour lines and $n!$
colour flows (i.e.\ there are $n!$ basis
tensors).  Inner products of colour flow basis tensors are given by
\begin{equation}
\langle \sigma|\tau\rangle =
\delta^{\alpha_1}_{\bar{\alpha}_{\sigma(1)}}\cdots
\delta^{\alpha_n}_{\bar{\alpha}_{\sigma(n)}}
\delta^{\bar{\alpha}_{\tau(1)}}_{\alpha_1}\cdots
\delta^{\bar{\alpha}_{\tau(n)}}_{\alpha_n}
= N^{n-\#\text{transpositions}(\sigma,\tau)}~, \label{eq:trans}
\end{equation}
where $\#\text{transpositions}(\sigma,\tau)$ is the number of
transpositions by which the permutations $\sigma$ and $\tau$
differ. This is equal to $n$ minus the number of loops obtained after
contracting the Kronecker symbols, see the right-hand part of
Figure~\ref{fig:colourflows}. In Figure~\ref{fig:colourflows}, we show
three of the six colour flows that represent the four-parton state on
the left, and in Table \ref{tab:flow} we specify the corresponding
colour and anti-colour indices for each of the four partons (labelled
by~$i$). We also include the binary variables $\lambda_i$ and
$\bar{\lambda}_i$, where $\lambda_i = \sqrt{T_R},\bar{\lambda}_{i}=0$
for a quark, $\lambda_i = 0,\bar{\lambda}_{i}=\sqrt{T_R}$ for an
antiquark and $\lambda_i = \bar{\lambda}_{i} = \sqrt{T_R}$ for a gluon
($T_R = 1/2$ in QCD). Note that in the figure we use the more compact
notation:
\begin{align}
|213 \rangle =\left|
\begin{array}{ccc}
1 & 2 & 3 \\
2 & 1 & 3
\end{array}  \right\rangle   ~~~~\text{etc.}
\end{align}

\begin{figure}
\begin{center}
\includegraphics[scale=0.6]{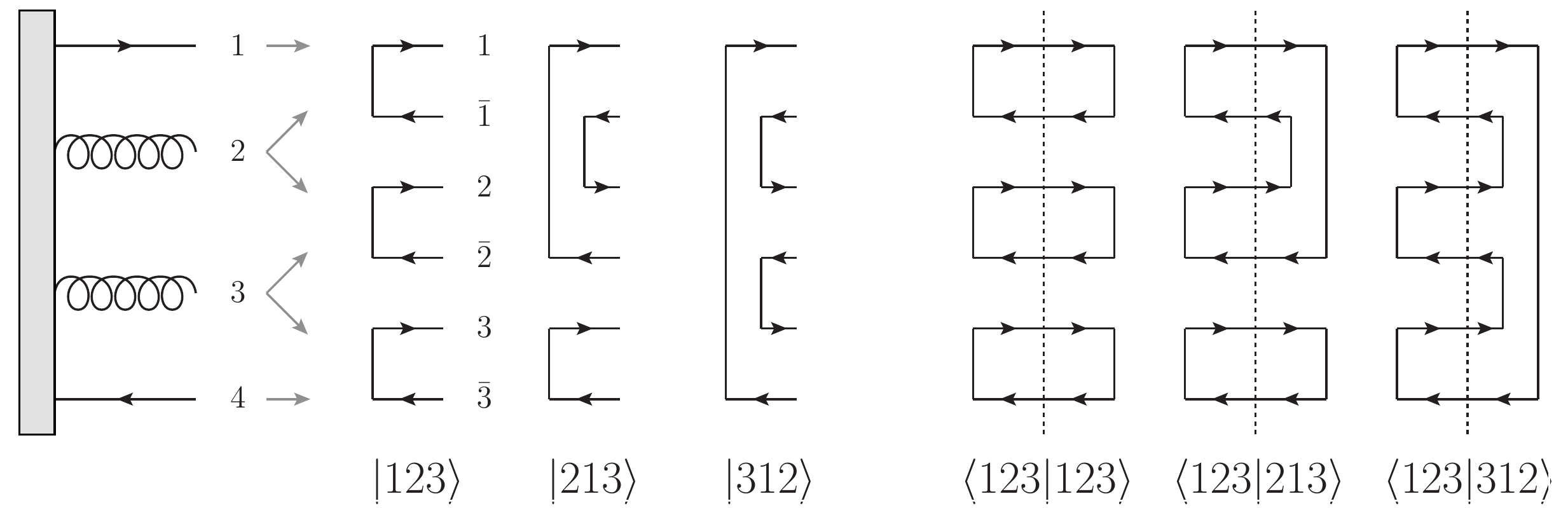}
\end{center}
\caption{\label{fig:colourflows}Diagrammatic representation of colour
  flow basis states and their inner products. The left half of the
  figure shows three out of the six basis tensors required for a
  $qgg\bar{q}$ leg content, where the grey arrows indicate how leg
  labels $i=1,..,4$ are mapped onto colour and anti-colour indices
  (see also Table \ref{tab:flow}). The right hand part of the figure
  illustrates how inner products, i.e.\ elements of the scalar product
  matrix, relate to powers of $N$ depending on how many loops are
  formed after contraction (see Eq.~(\ref{eq:trans})).}
\end{figure}

\begin{table}
\begin{center}
\vspace{0.5cm}
  \begin{tabular}{ | c | c | c | c | c | }
    \hline
    $i$ & $c_i$ & $\bar{c}_i$ & $\lambda_i$ & $\bar{\lambda}_i$ \\ \hline
    1 & 1 & 0 & $\sqrt{T_R}$ & 0 \\
    2 & 2 & 1 & $\sqrt{T_R}$ & $\sqrt{T_R}$ \\
   3 & 3 & 2 & $\sqrt{T_R}$ & $\sqrt{T_R}$ \\
   4 & 0 & 3 & 0 & $\sqrt{T_R}$ \\
    \hline
  \end{tabular}
\caption{The colour index specifications for the four partons in
  Figure \ref{fig:colourflows}, which correspond to $n =
  3$. \label{tab:flow}}.
\end{center}
\end{table}

We express amplitudes as $|{\cal A}\rangle = \sum_\sigma {\cal A}_\sigma
|\sigma\rangle$, where $\sigma$ labels the individual basis tensors, and the
evolution and traces in colour space can be performed in terms of ordinary
complex matrices with elements ${\cal A}_{\tau\sigma}$, which relate to the
basis independent objects via
\begin{equation}
{\mathbf A} = \sum_{\tau,\sigma} {\cal A}_{\tau\sigma}|\tau\rangle\langle
\sigma| \ .
\end{equation}
The coefficients ${\cal A}_{\tau\sigma}$ are not matrix elements of
the operator $\mathbf{A}$ since the colour flow basis is not
orthonormal. Consequently, we will introduce a dual basis in which
\begin{equation}
 [\tau|{\mathbf A}|\sigma] \equiv {\cal A}_{\tau\sigma}~.
\end{equation}
We refer to Appendix \ref{sec:dual} for more details on the properties of the
dual basis vectors (see also \cite{Platzer:2013fha}). The scalar product
matrix $S_{\tau\sigma} = \langle \tau|\sigma\rangle $ has to be considered
when evaluating the traces of operators in colour space:
\begin{equation}
{\rm Tr} [{\mathbf A}] = {\rm Tr} [{\cal A} S] = \sum_{\tau,\sigma}
[\tau|{\mathbf A}|\sigma]\ \langle \sigma| \tau\rangle \ .
\end{equation}

The colour charge (or emission) operator associated to each leg $i$ can be
decomposed as
\begin{equation}
{\mathbf T}_i = \lambda_i\ {\mathbf t}_{c_i} - \bar{\lambda}_{i}\
  \bar{{\mathbf t}}_{\bar{c}_i} - \frac{1}{N}(\lambda_i -
  \bar{\lambda}_{i})\ {\mathbf s}  \ ,
\end{equation}
where the colour-line operators
${\mathbf t},\bar{{\mathbf t}}$ and ${\mathbf s}$ are
defined through their action on the basis states, i.e.
\begin{equation}
{\mathbf t}_\alpha |\sigma\rangle ={\mathbf t}_\alpha \left|
\begin{array}{ccccc}
1 & \cdots & \alpha & \cdots & n \\
\sigma(1) & \cdots & \sigma(\alpha) &\cdots & \sigma(n)
\end{array}
\right\rangle =
\left|
\begin{array}{cccccc}
1 & \cdots & \alpha & \cdots & n & n+1 \\
\sigma(1) & \cdots & n+1 &\cdots & \sigma(n) & \sigma(\alpha)
\end{array}
\right\rangle \ ,
\end{equation}
\begin{equation}
\bar{{\mathbf t}}_{\bar{\alpha}}|\sigma\rangle = {\mathbf
  t}_{\sigma^{-1}(\bar{\alpha})}|\sigma\rangle\ ,
\end{equation}
for the inverse permutation $\sigma^{-1}$ for which
$\alpha=\sigma^{-1}(\sigma(\alpha))$, and
\begin{equation}
{\mathbf s} |\sigma\rangle ={\mathbf s} \left|
\begin{array}{cccc}
1 & \cdots & \cdots & n \\ \sigma(1) & \cdots&\cdots & \sigma(n)
\end{array}
\right\rangle =
\left|
\begin{array}{ccccc}
1 & \cdots &  \cdots & n & n+1 \\
\sigma(1) & \cdots &\cdots & \sigma(n) & n+1
\end{array}
\right\rangle \ .
\end{equation}
It is useful to note that
\begin{equation}
{\mathbf t}_\alpha|\sigma\rangle = {\mathbf s}_{\alpha,n+1} \, {\mathbf
  s} \, |\sigma\rangle \ ,
\end{equation}
where ${\mathbf s}_{\alpha,\beta}$ exchanges $\sigma(\alpha)$ and
$\sigma(\beta)$.  It is hence obvious that through the action of any
of the emission operators we cannot map two distinct basis tensors
$|\sigma\rangle$ and $|\tau\rangle$ into the same tensor
$|\rho\rangle$. Furthermore, if $\sigma$ and $\tau$ differ by $n$
transpositions, then
\begin{itemize}
\item for $\alpha\ne \beta$ ${\mathbf t}_\alpha|\sigma\rangle$ and
  ${\mathbf t}_\beta|\tau\rangle$ will differ by $n+2$ transpositions
  if $\sigma(\alpha)\ne \tau(\beta)$, and by the original $n$
  transpositions if $\sigma(\alpha)=\tau(\beta)$ (implying in this
  case that $n\ge 1$),
\item ${\mathbf t}_\alpha|\sigma\rangle$ and ${\mathbf s}|\tau\rangle$
  will differ by $n+1$ transpositions,
\item ${\mathbf s}|\sigma\rangle$ and ${\mathbf s}|\tau\rangle$ will
  differ by the original $n$ transpositions.
\end{itemize}

Colour-line operators and their products, such as $\mathbf{t}_\alpha \cdot
\mathbf{t}_\beta = \mathbf{t}_\beta \cdot \mathbf{t}_\alpha $, are referred to
as {\it colour reconnectors} in \cite{Platzer:2013fha}. We note that
$\mathbf{s} \cdot \mathbf{t}_\alpha = \mathbf{t}_\alpha \cdot \mathbf{s} =
\mathbf{1}$ and $\mathbf{s} \cdot \mathbf{s} = N\mathbf{1}$. Matrix elements
involving colour reconnectors are straightforward to compute because of the
important property that \beq \mathbf{R} \, | \sigma \rangle = [\tau |
  \mathbf{R} | \sigma \rangle \; |\tau \rangle~, \eeq where $\mathbf{R}$ is a
  general reconnector (see Appendix \ref{sec:dual}). Note that there is no sum
  over $\tau$ since reconnectors constitute a unique map from one colour flow
  to another. The matrix elements of the colour correlators are
\begin{eqnarray}
 \label{eq:titj}
[\tau|{\mathbf T}_i\cdot {\mathbf T}_j|\sigma \rangle &=&
-N \delta_{\tau\sigma}\left(
\lambda_i\bar{\lambda}_j\ \delta_{c_i,\sigma^{-1}(\bar{c}_j)}
+
\lambda_j\bar{\lambda}_i\ \delta_{c_j,\sigma^{-1}(\bar{c}_i)}
+
\frac{1}{N^2}(\lambda_i - \bar{\lambda}_{i})(\lambda_j -
\bar{\lambda}_{j})\right)\nonumber \\
&& +\sum_{(ab)}\delta_{\tau_{(ab)},\sigma}\left( \lambda_i\lambda_j
\delta_{(ab),(c_i c_j)} + \bar{\lambda}_i\bar{\lambda}_j
\delta_{(ab),(\sigma^{-1}(\bar{c}_i) \sigma^{-1}(\bar{c}_j))}
\right.\\\nonumber&&\qquad\qquad\qquad \left.
-\lambda_i\bar{\lambda}_j
\delta_{(ab),(c_i,\sigma^{-1}(\bar{c}_j))} -
\lambda_j\bar{\lambda}_i
\delta_{(ab),(c_j,\sigma^{-1}(\bar{c}_i))}
 \right)~,
\end{eqnarray}
for $i\ne j$, where $(ab)$ denotes an ordered pair ($(ab)=ba$ if $a>b$) and
$\tau_{(ab)}$ denotes swapping the elements $a$ and $b$ in the permutation
$\tau$.  $\delta_{\tau\sigma}$ is zero if the permutations $\tau$ and $\sigma$
are not equal and unity otherwise. The sum over $(ab)$ is rather cumbersome,
since each of the four terms can be written without any summation after
implementing the colour-flow Kronecker delta. However, this way of writing
things ensures that the second and third line in Eq.~(\ref{eq:titj}) do not
contribute if $i$ and $j$ are colour connected in $\sigma$. By `colour
connected' we mean $c_i = \sigma^{-1}(\bar{c}_j)$ or $c_j =
\sigma^{-1}(\bar{c}_i)$.\footnote{If $\sigma = |321\rangle$ then $\sigma(3) =
  1$ and $\sigma^{-1}(3)=1$ etc.}  Note also that the off-diagonal elements in
the matrix representation of ${\mathbf T}_i\cdot {\mathbf T}_j$ are
non-vanishing only if the permutations labeling the two basis tensors in
question differ by at most one transposition.  A similar expression can be
obtained for colour charges multiplied to the left and right of a colour
matrix, $\mathbf{A}$, which corresponds to real emission:
\begin{eqnarray}
\label{eqn:emittors}
[\sigma|{\mathbf T}_i \, {\mathbf A} \, {\mathbf T}_j|\tau ]
= \Big\{ \Big( &-& \lambda_i\bar{\lambda}_j
\delta_{c_i\sigma^{-1}(\bar{c}_n)}\delta_{\bar{c}_j \tau(c_n)} -
\left(i,\sigma\leftrightarrow j,\tau\right)   \Big) \\\nonumber
&+& \lambda_i\lambda_j
\delta_{c_i\sigma^{-1}(\bar{c}_n)}\delta_{c_j\tau^{-1}(\bar{c}_n)} +
\bar{\lambda}_i\bar{\lambda}_j
\delta_{\bar{c}_i\sigma(c_n)}\delta_{\bar{c}_j\tau(c_n)}\\\nonumber
\Big( &-&
\frac{1}{N}\left(\lambda_i\delta_{c_i\sigma^{-1}(\bar{c}_n)}-
\bar{\lambda}_i\delta_{\bar{c}_i\sigma(c_n)}\right)(\lambda_j-\bar{\lambda}_j)\delta_{c_n\tau^{-1}(\bar{c}_n)}-
\left(i,\sigma\leftrightarrow j,\tau\right) \Big) \\\nonumber
&+&
\frac{1}{N^2}(\lambda_i-\bar{\lambda}_i)(\lambda_j-\bar{\lambda}_j)\delta_{c_n\sigma^{-1}(\bar{c}_n)}\delta_{c_n\tau^{-1}(\bar{c}_n)}\Big\}\\\nonumber
&\times& \left[\tau\backslash n|{\mathbf A}|\sigma\backslash n \right ]\ .
\end{eqnarray}
The colour lines associated with the emitted particle, $n$, are
labelled by $c_n$ and $\bar{c}_n$, and $\sigma\backslash n$
denotes the permutation with the entries associated with $c_n$ and
$\bar{c}_n$ merged and removed, i.e.
\begin{equation}
\left(
\begin{array}{ccccccc}
1 & \cdots & c_n & \cdots & \sigma^{-1}(\bar{c}_n) & \cdots & m \\
\sigma(1) & \cdots & \sigma(c_n) &\cdots& \bar{c}_n &\cdots & \sigma(m)
\end{array}\right)\backslash n = \left(
\begin{array}{ccccc}
1&\cdots &\sigma^{-1}(\bar{c}_n)& \cdots & m\\
\sigma(1)&\cdots & \sigma(c_n)&\cdots & \sigma(m)
\end{array}
\right) \ .
\end{equation}

Our aim is to organize contributions to the cross section in terms of a
series of leading powers in $N$, to extract both the large-$N$ limit as well as
corrections to it. To this end we introduce the operation
\begin{equation}
\text{Leading}_{\tau\sigma}^{(l)}\left[{\mathbf A}\right] = \sum_{k=0}^l
 {\cal A}_{\tau\sigma}\Big|_{1/N^{k}}\
\delta_{\#\text{transpositions}(\tau,\sigma),l-k}
\ ,
\end{equation}
where the notation
\begin{equation}
 {\cal A}_{\tau\sigma}\Big|_{1/N^{k}}
\end{equation}
indicates to pick those terms in ${\cal A}_{\tau\sigma}$ which are
suppressed by a factor of $1/N^{k}$ with respect to the leading power
present in ${\cal A}_{\tau\sigma}$. Contributions to the trace of
${\mathbf A}$ then all yield an enhancement or a supression by the
same power of $N$ by virtue of either an explicit suppression in
${\cal A}_{\tau\sigma}$ or by picking up a subleading element in the
scalar product matrix. In other words, if ${\mathbf A}\sim {\mathbf
  A}_0 + {\mathbf A}_1/N+...$ is an operator in the space of $n$
colour lines, then
\begin{equation}
{\rm Tr}\left[\text{Leading}^{(l)}\left[{\mathbf A}\right] \right]\propto
N^{n-l} \ .
\end{equation}
We are specifically interested in traces originating from soft-gluon evolution:
\begin{equation}
{\rm Tr}[{\mathbf V}_n {\mathbf A}_n {\mathbf V}_n^\dagger]\qquad \text{where} \qquad
{\mathbf V}_n = \exp\left( \sum_{i,j} \Omega^{(n)}_{ij}
{\mathbf T}_i\cdot {\mathbf T}_j\right) \ ,
\end{equation}
\begin{equation}
{\rm Tr}[\boldsymbol{\gamma}_n {\mathbf A}_{n} + {\mathbf
A}_{n}\boldsymbol{\gamma}_n^\dagger] \qquad \text{where} \qquad
\boldsymbol{\gamma}_n = \sum_{i,j}\gamma^{(n)}_{ij}{\mathbf T}_i\cdot
      {\mathbf T}_j \ ,
\end{equation}
and
\begin{equation}
\sum_\lambda {\rm Tr}\Big[{\mathbf D}^\lambda_n {\mathbf A}_{n-1} \left({\mathbf D}_n^\lambda\right)^\dagger\Big]\qquad \text{where} \qquad
{\mathbf D}^\lambda_n = \sum_{i} \omega^{\lambda,(n)}_i {\mathbf T}_i \ ,
\end{equation}
where $\lambda$ refers to all other quantum numbers of the emission, and it is understood that
$\sum_\lambda\omega^{\lambda,(n)}_i (\omega^{\lambda,(n)}_i)^* = 0$.

\subsection{Leading contributions} \label{leadingN}

For the leading contributions of the virtual evolution operators we find
\begin{equation}
\text{Leading}_{\tau\sigma}^{(0)}\left[{\mathbf V}_n {\mathbf A}_n {\mathbf
    V}_n^\dagger\right] =
\delta_{\tau\sigma}\left| V_\sigma^{(n)}\right|^2
\ \text{Leading}_{\tau\sigma}^{(0)}\left[{\mathbf A}_n\right] \ ,
\end{equation}
where
\begin{equation}
V_\sigma^{(n)} = \exp\left(-N\sum_{i,j\text{ c.c. in }\sigma} \lambda_i
\bar{\lambda}_j W_{ij}^{(n)}
\right)
\end{equation}
and c.c. here means colour connected, i.e.
\begin{equation}
\sum_{i,j\text{ c.c. in }\sigma} x_{ij} =
\sum_{i,j} x_{ij}\ \delta_{c_i\sigma^{-1}(\bar{c}_j)}~.
\end{equation}
Also, we have defined
\begin{equation}
W_{ij}^{(n)} = \left(\Omega^{(n)}_{ij}+\Omega^{(n)}_{ji}\right)\ .
\end{equation}
For single gluon exchange,
\begin{equation}
\text{Leading}_{\tau\sigma}^{(0)}\left[\boldsymbol{\gamma}_n {\mathbf
    A}_n + {\mathbf A}_n\boldsymbol{\gamma}_n^\dagger\right] = -N\
\delta_{\tau\sigma}\sum_{i,j\text{ c.c. in }\sigma}
\lambda_{i}\bar{\lambda}_j r_{ij}^{(n)}\ \text{Leading}_{\tau\sigma}^{(0)}\left[{\mathbf A}_n\right]
\ ,
\end{equation}
with
\begin{equation}
r_{ij}^{(n)} = \ 2{\rm Re}\left( \gamma_{ij}^{(n)} +
\gamma_{ji}^{(n)}\right) \ .
\end{equation}
And the emission contribution is
\begin{equation}
\text{Leading}_{\tau\sigma}^{(0)}\left[{\mathbf D}_n {\mathbf A}_{n-1} {\mathbf
    D}^\dagger_n\right] =\delta_{\tau\sigma}\sum_{i,j\text{ c.c. in
  }\sigma\backslash n}\lambda_{i}\bar{\lambda}_j R_{ij}^{(n)}\ \text{Leading}_{\tau\backslash
  n,\sigma\backslash n}^{(0)}\left[{\mathbf A}_{n-1}\right]
\end{equation}
where
\begin{equation}
R_{ij}^{(n)} = -
\ 2{\rm Re}\left( \omega^{(n)}_i \left(\omega^{(n)}_j\right)^*\right) \ .
\end{equation}
We are now able to compute traces in the leading-$N$ limit. We must
sum over diagonal colour flow contributions,
i.e.\ $A_{\sigma\sigma}$. For each colour flow, $\sigma$, we multiply
by $N$ raised to the number of colour lines present in the colour flow
$\sigma$. Notice that the number of possible colour flows at this
level of approximation is equal to the number present at the level of the
hard process plus the number of real emissions.  Each of the
contributions $A_{\sigma\sigma}$ can then be computed by a set of
recursive rules that correspond to working inwards from the outer
matrices (multiplied from the left and right) towards the hard matrix
in between.
\begin{figure}
\begin{center}
\includegraphics[scale=0.6]{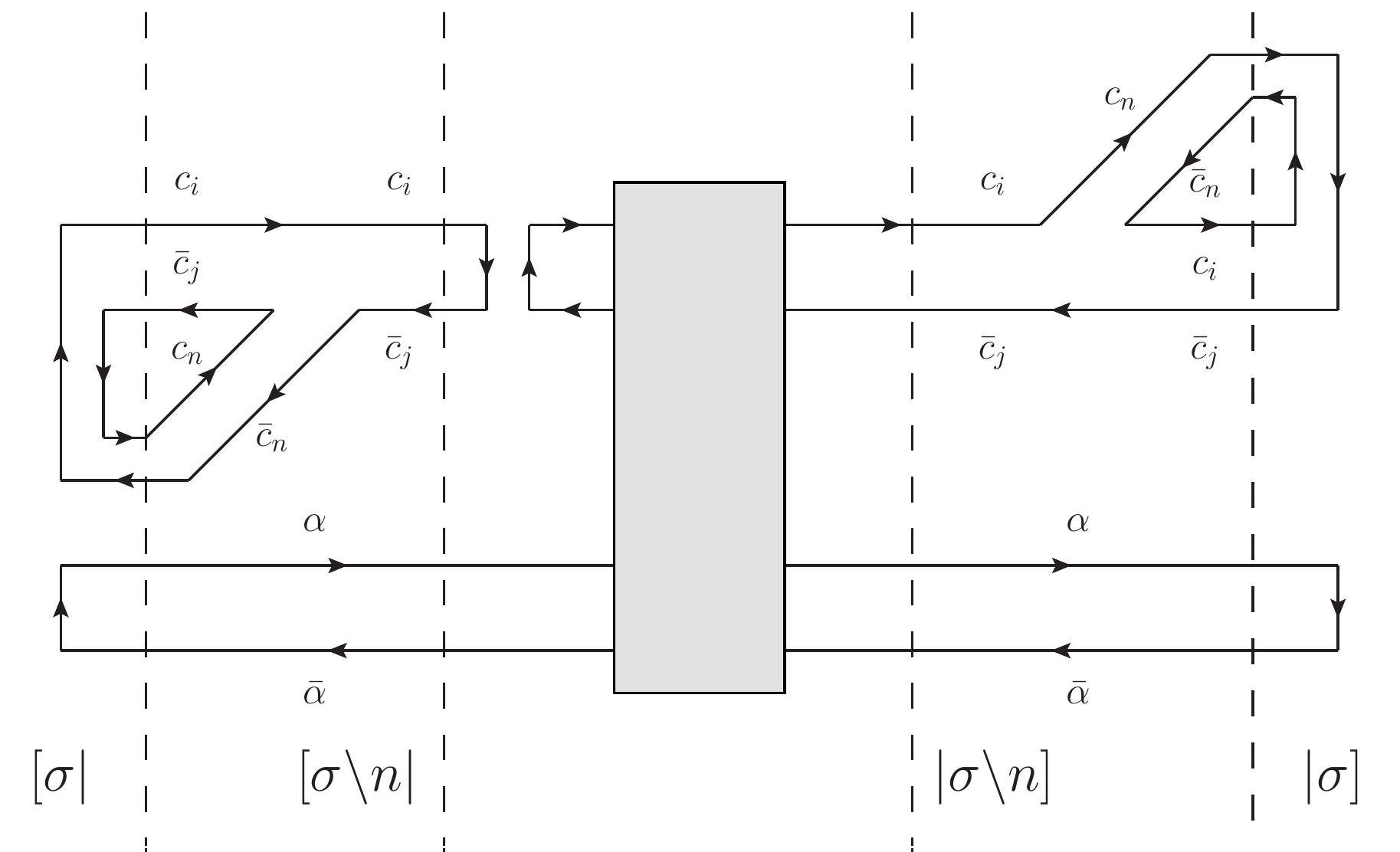}
\end{center}
\caption{\label{figures:dipolemerging} An example to illustrate
  colour-flow evolution. In this case, we consider the leading-colour
  contribution, with one real emission and one virtual correction, to
  $[ \tau | \mathbf{A} | \sigma]$. The vertical dashed lines indicate
  that it is possible to read off the colour-flow map at any time
  during the evolution. See the text for the corresponding rules. }
\end{figure}
The rules are as follows:
\begin{itemize}
\item For a pair of evolution operators, ${\mathbf V}_n {\mathbf A}_n {\mathbf
  V}^\dagger_n$, multiply by
\begin{align}
\exp\left(-2N\ {\rm Re}\left( \Omega_{ij}^{(n)} + \Omega_{ji}^{(n)}\right) \lambda_i\bar{\lambda}_j\right)
\end{align} for each colour connected dipole $i,j$ in $\sigma$.
\item For a virtual gluon insertion, $\boldsymbol{\gamma}_n{\mathbf A}_n +
  {\mathbf A}_n \boldsymbol{\gamma}^\dagger_n$, multiply by
\begin{align}
-2N\ \lambda_i \bar{\lambda}_j\ {\rm Re}\left(\gamma^{(n)}_{ij} + \gamma^{(n)}_{ji}\right)
\end{align}
and sum over the dipoles $(i,j)$ that are colour connected in $\sigma$.
\item For a pair of emission operators, ${\mathbf D}_n{\mathbf A}_{n-1}
{\mathbf
  D}_n^\dagger$, combine the dipoles $(i,n)$ and $(n,j)$ in the colour flow $\sigma$, leaving behind a dipole $(i,j)$ in the colour flow $\sigma\backslash n$, and include a factor
\begin{align} -\lambda_i \bar{\lambda}_j\ 2 \,{\rm Re}\left(\omega_i^{(n)}\left(\omega_j^{(n)}\right)^*\right) \ .
\end{align} This procedure is illustrated in
Figure~\ref{figures:dipolemerging}, where we take the opportunity to
show a specific contribution at leading-colour. This graph would
contribute to the soft gluon evolution of $qq \to qq$ scattering with
one emission and one virtual correction.
\item If the hard process has been reached, multiply by
  the square of the corresponding amplitude, $|{\cal M}_\sigma|^2$.
\end{itemize}

\subsection{Dipole evolution and the BMS equation}
\label{sections:BMS}

We will now show how the rules of the preceding section give rise to
the BMS equation \cite{Banfi:2002hw}. Our rules apply to a general
process with any number of outgoing partons. The algorithmic
incarnation of the generalized BMS equation that we present here
corresponds to a dipole shower algorithm. The evolution of dipoles is
universal, i.e.\ at this level of approximation the process dependence
solely enters through selecting an initial colour flow weighted by the
modulus squared of the corresponding amplitude $|{\cal
  M}_\sigma|^2$. To illustrate how things work out, we will consider
the same example as in Section \ref{sec:examples}. In this case
\begin{equation}
\Omega_{ij}^{(n)} = \Omega_{ji}^{(n)} = -\int_\text{in} \dd \Pi_k \,  \Theta(E_{n+1}< E < E_{n}) \,
\frac{1}{2}\omega_{ij}(\hat{k}) \ ,
\end{equation}
\begin{equation}
\gamma_{ij}^{(n)} = \gamma_{ji}^{(n)} = - \Theta_\text{out} \, \Theta(E_n < E_{n-1}) \,  \frac{1}{2}\omega_{ij}(\hat{q}_n) \ ,
\end{equation}
and
\begin{equation}
\omega^{(n)}_i\left(\omega^{(n)}_j\right)^* = \Theta_\text{out} \, \Theta(E_n < E_{n-1}) \, \omega_{ij}(\hat{q}_n)~.
\end{equation}
The
evolution with the in-region anomalous dimension contributes a factor
\begin{equation}
V_{ij}^{E_{n+1},E_n} = \exp\left(-\frac{N\alpha_s}{\pi} \int_{E_{n+1}}^{E_n} \frac{{\rm d}E}{E} \int_\text{in} \frac{{\rm
    d}\Omega_k}{4\pi} \omega_{ij}(\hat{k}) \right)
\end{equation}
per colour flow.

\begin{figure}[t]
\centering
  \includegraphics[width=0.8\textwidth]{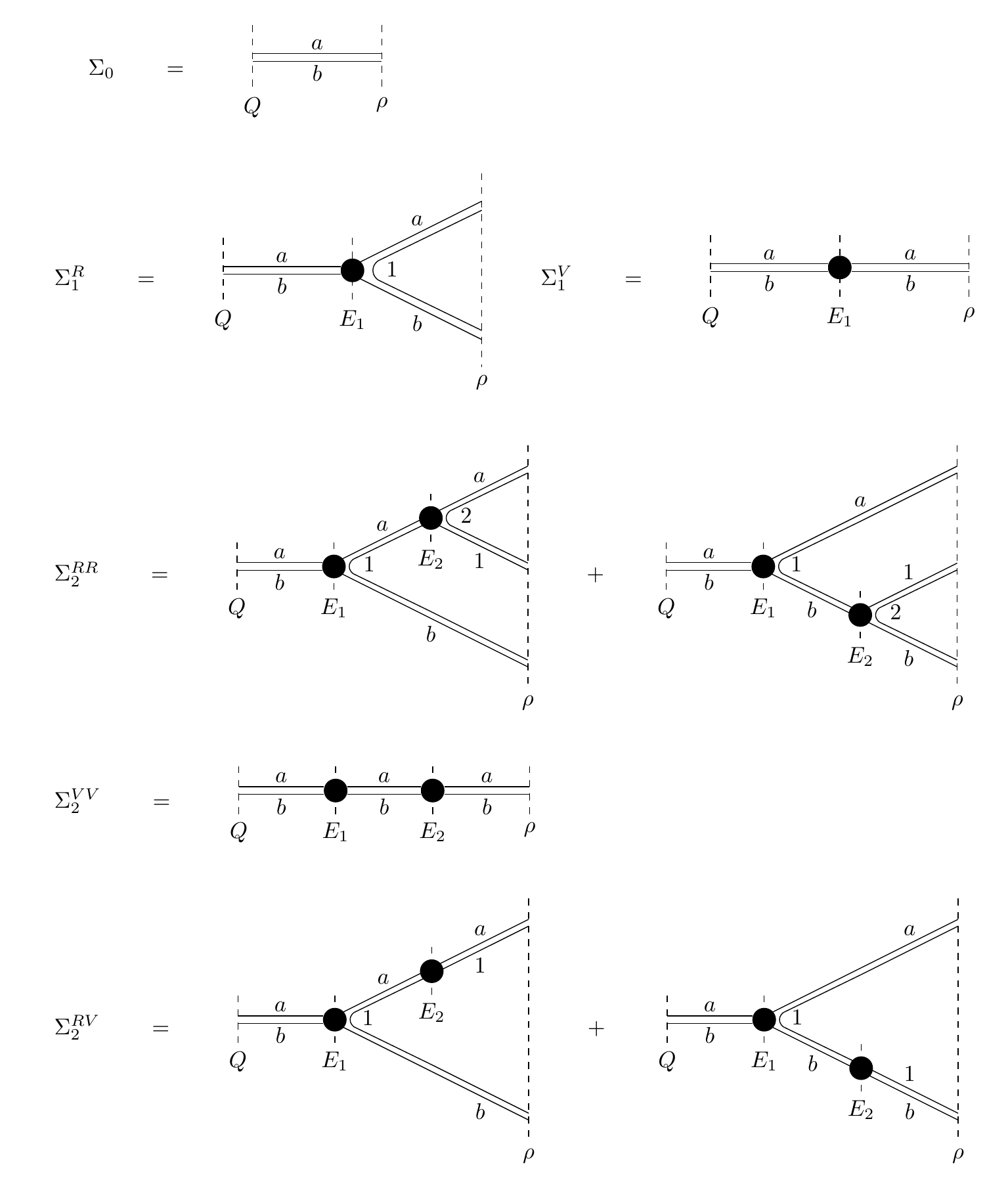}
  \caption{The leading-$N$ graphs corresponding to Eq.~(\ref{eq:sigmaN})}. \label{fig:dipole}
\end{figure}

The above expressions have a very simple diagrammatic interpretation,
illustrated in Figure \ref{fig:dipole}. To simplify the discussion, we
consider the case of $e^+e^-$ scattering, i.e.\ we take $\mathbf{H} =
\frac{1}{N}\mathbf{1}$. Each double line in the figure corresponds to
a Sudakov factor, $V_{ij}^{E_1,E_2}$, where $i$ and $j$ label the
directions associated with the corresponding colour and anti-colour
lines. The shaded circles correspond to a factor
$\omega_{ij}(\hat{k})$, and the vertical dashed line indicates the
associated energy. The arguments of the Sudakov are also determined by
these vertical dashed lines. We can immediately see how the algorithm
maps onto a classical dipole shower at leading $N$.  These
diagrammatic rules can be used to compute the leading colour
contribution to the non-global logarithms: \bea \Sigma_0 &=&
V_{ab}^{\rho,Q} \nonumber \\ \Sigma_1 &=& \int_\text{out}
V_{ab}^{E_1,Q} \omega_{ab}^1 \; \Big[ V_{a1}^{\rho,E_1}
  V_{b1}^{\rho,E_1} - V_{ab}^{\rho,E_1} \Big] ~ \frac{\dd
  \Omega_1}{4\pi} \dd t_1 \nonumber \\ \Sigma_2^{RR} &=&
\int_\text{out} V_{ab}^{E_1,Q} \omega_{ab}^1 \Big[ V_{a1}^{E_2,E_1}
  V_{b1}^{\rho,E_1} \omega_{a1}^2 V_{a2}^{\rho,E_2} +
  V_{a1}^{\rho,E_1} V_{b1}^{E_2,E_1} \omega_{b1}^2 V_{b2}^{\rho,E_2}
  \Big] V_{12}^{\rho,E_2} \nonumber \\ & & \times \frac{\dd
  \Omega_1}{4\pi} \frac{\dd \Omega_2}{4\pi} \dd t_1 \dd t_2 \nonumber
\\ \Sigma_2^{VR} &=& - \int_\text{out} V_{ab}^{E_2,Q} \omega_{ab}^1
\omega_{ab}^2 V_{a1}^{\rho,E_2} V_{b1}^{\rho,E_2} \frac{\dd
  \Omega_1}{4\pi} \frac{\dd \Omega_2}{4\pi} \dd t_1 \dd t_2 \nonumber
\\ \nonumber \\ \Sigma_2^{RV} &=& -\int_\text{out} V_{ab}^{E_1,Q}
\omega_{ab}^1 \Big[ V_{a1}^{E_2,E_1} V_{b1}^{\rho,E_1} \omega_{a1}^2
  V_{a1}^{\rho,E_2} + V_{a1}^{\rho,E_1} V_{b1}^{E_2,E_1} \omega_{b1}^2
  V_{b1}^{\rho,E_2} \Big] \nonumber \\ & & \times \frac{\dd
  \Omega_1}{4\pi} \frac{\dd \Omega_2}{4\pi} \dd t_1 \dd t_2 \nonumber
\\ \Sigma_2^{VV} &=& \int_\text{out} V_{ab}^{\rho,Q} \omega_{ab}^1
\omega_{ab}^2 \frac{\dd \Omega_1}{4\pi} \frac{\dd \Omega_2}{4\pi} \dd
t_1 \dd t_2 \nonumber \\ & \mathrm{ etc.} & \label{eq:sigmaN} \eea
where the hard partons have momenta $p_a$ and $p_b$, $t_i = (N
\alpha_s /\pi) \ln (E_i/\rho)$ and we used the notation $\omega_{ab}^i
= \omega_{ab}(\hat{q}_i)$.

The $\Sigma_n$ can also be obtained by iteratively solving the BMS
equation, as we will now illustrate. The BMS equation can be written
as follows, \beq \frac{\partial G_{ab}(t)}{\partial t} =
-\int_{\text{in}} \frac{\dd\Omega_k}{4\pi} \omega_{ab}(k) G_{ab}(t) +
\int_{\text{out}}\frac{\dd\Omega_k}{4\pi} \omega_{ab}(k) \Big[
  G_{ak}(t) G_{kb}(t) - G_{ab}(t) \Big]~ \eeq and our observable
corresponds to \beq \Sigma(\rho) = G_{ab}(t_Q) \eeq with $G_{ab}(0) =
1$ and $t_Q = (N \alpha_s /\pi) \ln (Q/\rho)$.  To solve the BMS
equation iteratively, we will first rewrite it by replacing $G_{ij}(t)
= V_{ij}^{\rho,E} g_{ij}(t)$, which gives \beq \frac{\partial
  g_{ab}(t)}{\partial t} = \int_{\text{out}}\frac{\dd\Omega_k}{4\pi}
\omega_{ab}(k) \left[ \frac{V_{ak}^{E,\rho}
    V_{kb}^{E,\rho}}{V_{ab}^{E,\rho}} \; g_{ak}(t) g_{kb}(t) -
  g_{ab}(t) \right]~. \label{eq:BMS} \eeq Putting $g_{ab}^{(0)}(t)=1$
on the RHS of the BMS equation immediately gives $\Sigma_1$, i.e.
\beq \frac{\partial g_{ab}^{(1)}(t)}{\partial t} =
\int_{\text{out}}\frac{\dd\Omega_k}{4\pi} \omega_{ab}(k) \left[
  \frac{V_{ak}^{E,\rho} V_{kb}^{E,\rho}}{V_{ab}^{E,\rho}} - 1
  \right]~, \eeq which gives the desired result after integrating over
$0 < t < t_Q$.  The next iteration gives $\Sigma_2$, i.e.\ we
substitute $g_{ij}(t)$ on the RHS of the BMS equation by
$g_{ab}^{(1)}(t)$: \beq \frac{\partial g_{ab}^{(2)}(t)}{\partial t} =
\int_{\text{out}}\frac{\dd\Omega_k}{4\pi} \omega_{ab}(k) \left[
  \frac{V_{ak}^{E,\rho} V_{kb}^{E,\rho}}{V_{ab}^{E,\rho}} \;
  \left(g_{ak}^{(1)}(t) g_{kb}^{(0)}(t) +g_{ak}^{(0)}(t)
  g_{kb}^{(1)}(t)\right) - g_{ab}^{(1)}(t) \right]~, \eeq where we
left $g_{ak}^{(0)}=1$ explicit for clarity.  It is easy to show
that $V_{ab}^{Q,\rho} g_{ab}^{(2)}(t_Q) = \Sigma_2^{RR} +
\Sigma_2^{VR}+\Sigma_2^{RV}+\Sigma_2^{VV}$. So we see that, at leading
$N$, our algorithm generates the iterative solution to the BMS
equation.

\subsection{Subleading contributions}

\begin{figure}
\begin{center}
\includegraphics[scale=0.5]{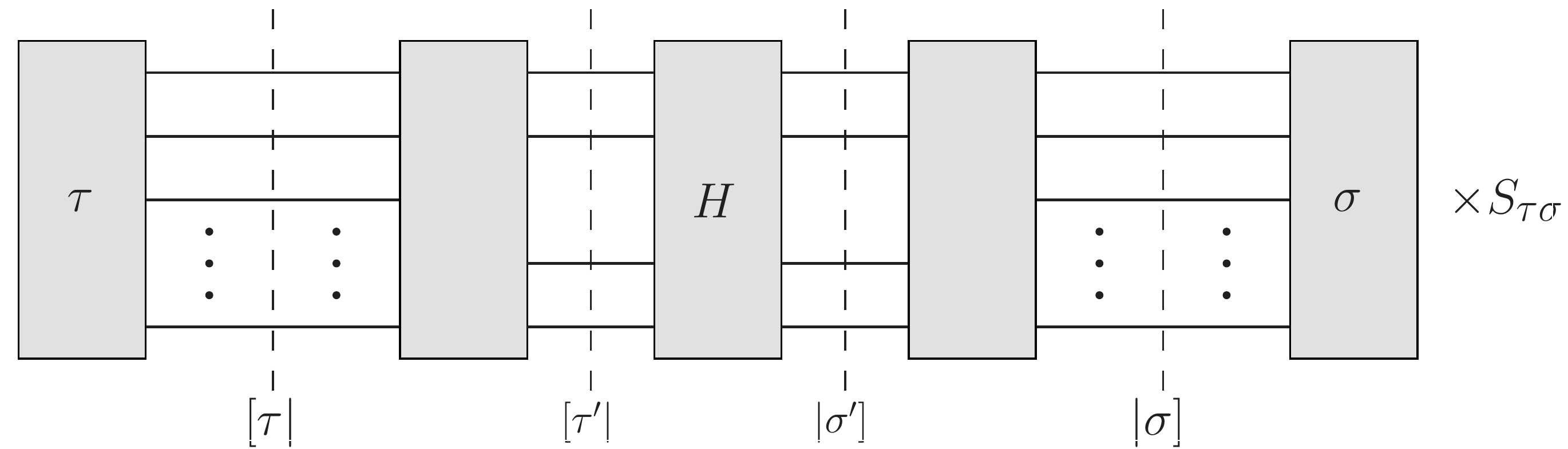}
\end{center}
\caption{\label{fig:generalstructure}The general structure of the
  calculation of subleading contributions.}
\end{figure}

Subleading colour contributions are substantially more difficult to
compute. In this section we present some initial steps towards a systematic
approach to including $1/N^k$ corrections to the leading
result. Figure~\ref{fig:generalstructure} illustrates the general structure of
the calculation (of which Figure~\ref{figures:dipolemerging} is a specific
example). Figure~\ref{figures:towers} provides an overview of the power
counting we use to define successive orders -- we hope its interpretation will
become clear after the following paragraphs.

There are subleading colour contributions arising from the hard
scattering matrix, from the $1/N$ and $1/N^2$ suppressed terms in the real
emission operator (see Eq.~(\ref{eqn:emittors})) and the virtual
correction operator (see Eq.~(\ref{eq:titj})), and from off-diagonal
contributions to the scalar product matrix. In the following, we will use
the general form of the anomalous dimension resulting from
Eq.~(\ref{eq:titj}), i.e.
\begin{equation}
[\tau| {\mathbf \Gamma} |\sigma \rangle = N \delta_{\tau\sigma}\Gamma_\sigma
+ \Sigma_{\tau\sigma} + \frac{1}{N}\delta_{\tau\sigma}\rho \ .
\end{equation}
Each of $\Gamma$, $\Sigma$ and $\rho$ are of order $\alpha_s$.
To compute a correction of order $1/N^k$ we need to consider
states $\sigma$ and $\tau$ in Figure \ref{fig:generalstructure} that
differ by $k-l$ permutations, where $0 \le l \le k$. Then we must
determine the $1/N^l$ corrections arising from the soft gluon
evolution and from the hard scattering matrix.

\begin{figure}
\begin{center}
\includegraphics[scale=0.5]{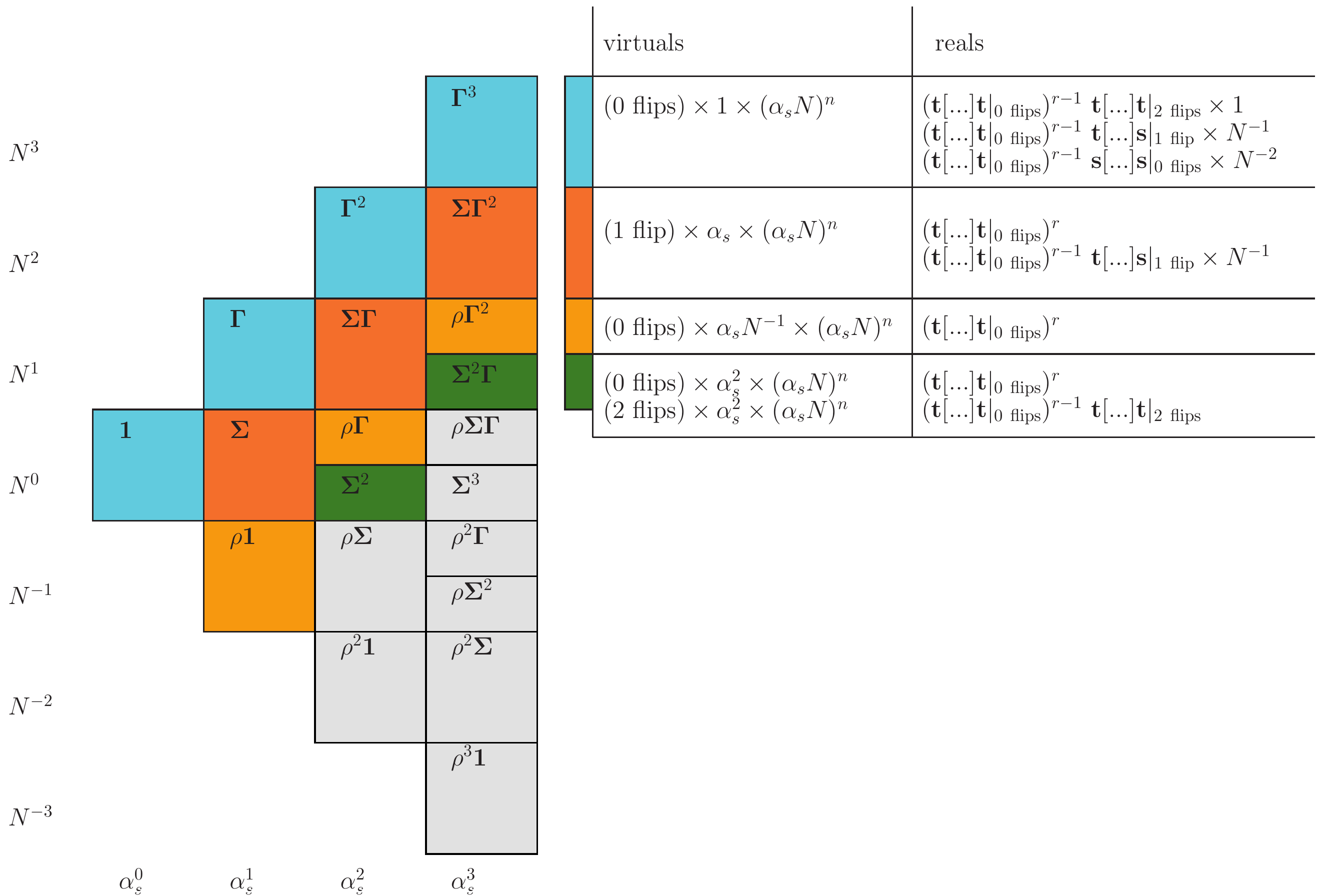}
\end{center}
\caption{\label{figures:towers}The counting of next-to-leading colour
  contributions. We aim to perform a resummation in all powers of the t'Hooft
  coupling $\alpha_s N\sim 1$. In the set of boxes shown we count, for the
  virtual evolution operator, increasing powers of $\alpha_s$ from left to
  right, and decreasing powers of $N$ from top to bottom, with $N^0$ in the
  middle row. The effect of $r$ real emissions is indicated in the rightmost
  column of the figure and any $1/N$ suppression due to the scalar product
  matrix is indicated by the number of flips.  See the text for more details.}
\end{figure}

The leading colour contributions from the virtual evolution operator come from
$\Gamma$ and so are all enhanced by powers of $\alpha_s N$ which, owing to the
fact that the leading contribution is diagonal, can easily be accounted for to
all orders in a simple exponential. This evolution does not result in any
difference between the colour structure in the amplitude and that in its
conjugate, and it corresponds to the blue boxes in
Figure~\ref{figures:towers}. If this evolution is then supplemented by those
pieces of the real emission operator that also preserve the identity of the
colour structure in the amplitude and its conjugate (such as the example in
Figure~\ref{figures:dipolemerging}) then we recover the leading-$N$ picture of
the last two sections.

Subleading colour contributions may result in differences between the colour
in the amplitude and that in the conjugate amplitude. To keep track of this,
we will count the number of colour reconnections (or transpositions or flips
or swings) by which the two colour structures differ. It turns out that pure
$1/N$ corrections can only originate from interference contributions in the
hard process matrix. We will ignore subleading colour contributions from this
source in what follows, though they could easily be included. The most important
subleading colour contributions due to real emission are suppressed by a power
of $1/N^2$ relative to the leading contribution and they originate as a result
of the following three possibilities: (i) two colour flips accompanied by no
explicit factor of $1/N$ (coming from contributions of the type $\mathbf{t} [
  \cdots ] \mathbf{t}$); (ii) one flip and a factor of $1/N$ (coming from
contributions of the type $\mathbf{t} [ \cdots ] \mathbf{s}$ and $\mathbf{s} [
  \cdots ] \mathbf{t}$); (iii) zero flips and a factor of $1/N^2$ (coming from
contributions of the type $\mathbf{s} [ \cdots ] \mathbf{s}$). See
Eq.~(\ref{eqn:emittors}) to appreciate the factors of $1/N$. We note that real
emissions never reduce the number of flips by which the amplitude and its
conjugate differ. We will present the explicit rules
corresponding to these real emission contributions below but first we consider subleading
virtual corrections.

A single insertion of a perturbation $\Sigma_{\sigma'\sigma}$ comes with a
factor of $(\alpha_s N)/N$ relative to the leading contribution and it results
in a single flip, and hence an additional $1/N$ suppression via the scalar
product matrix. This flip can undo that induced by a previous real emission of
the type $\mathbf{s}[ \cdots] \mathbf{t}$ or $\mathbf{t}[ \cdots]
\mathbf{s}$. However, since this will require the action of a single ${\mathbf
  s}$ operator, it re-introduces the additional factor of $1/N$. Thus, both of
these fixed-order contributions, when combined with the all-order summation of
contributions from $\Gamma$, are suppressed by $(\alpha_s N)/N^2\sim 1/N^2$
relative to the leading contributions. These contributions are illustrated by
the dark orange boxes in Figure~\ref{figures:towers}. A similar reasoning
applies to the contribution of a single $\rho$ perturbation (light orange
boxes), which contributes at the same order $\alpha_s/N \sim 1/N^2$ since it
generates zero flips.

We finally need to consider two insertions of $\Sigma_{\sigma'\sigma}$
combined in such a way that the net number of flips is zero or two. The
zero-flip case is clearly proportional to $\alpha_s^2$ and hence contributes a
$(\alpha_sN)^2/N^2$ correction (green boxes). The two-flip case can also
contribute at this order provided it compensates a $\mathbf{t}[ \cdots]
\mathbf{t}$ two-flip real emission, i.e.\ so the net result is that the
amplitude and its conjugate differ by zero flips and there is no suppression
from the scalar product matrix.  These two contributions should therefore be
included along with the contributions discussed above. However, contributions
from the diagonal below the green boxes lead to a factor of $(\alpha_s
N)^2/N^4$, which means they are beyond the next-to-leading colour
approximation.

These corrections to soft-gluon evolution can be considered as fixed-order
corrections to the leading-$N$ rules, though these need to be extended
to include the possibility that the permutations $\sigma$ and $\tau$
need no longer be equal. This means we should update the rules at the
end of Section \ref{leadingN} as follows.
\begin{itemize}
\item For a pair of evolution operators ${\mathbf V}_n {\mathbf A}_n {\mathbf
  V}^\dagger_n$ multiply by $V_{\sigma}^{(n)}\left(V_{\tau}^{(n)}\right)^*$,
  i.e.\ include a factor of the virtual amplitude exponentiated for each
  colour connected pair of legs in $\sigma$, and each colour connected pair in
  $\tau$.
\item For a virtual gluon insertion $\boldsymbol{\gamma}_n{\mathbf A}_n +
  {\mathbf A}_n \boldsymbol{\gamma}^\dagger_n$, include a factor
  $w_\sigma^{(n)} + \left(w_\tau^{(n)}\right)^*$, where
\begin{align}
w_\sigma^{(n)} = \sum_{i,j\text{ c.c. in }\sigma} \lambda_i\bar{\lambda}_j
\left(\gamma_{ij}^{(n)}+\gamma_{ji}^{(n)}\right) \ .
\end{align}
\item Real emission operators, ${\mathbf
  D}_n{\mathbf A}_{n-1} {\mathbf D}_n^\dagger$, only contribute if the emitted
  gluon is connected to the same, identically connected, colour line in
  $\sigma$ and $\tau$, as otherwise the operation of merging the emitting
  dipoles would alter the number of flips by which $\sigma$ and
  $\tau$ differ.
\end{itemize}

We will now present the rules to compute the first $1/N^2$ corrections to the
leading colour trace. For the virtual contributions we need to consider the
next-to-leading colour approximation for the evolution operator, which (using
the notation in \cite{Platzer:2013fha}) is
\begin{equation} \label{eq:sigmaflip}
{\mathbf V}_n^{\text{LC}+\text{NLC}}|\sigma\rangle = V_\sigma^{(n)}
|\sigma\rangle -
\frac{1}{N}\sum_{\tau}
\delta_{\#\text{transpositions}(\tau,\sigma),1}\Sigma^{(n)}_{\sigma\tau}
|\tau\rangle \ ,
\end{equation}
where the colour flow transition matrix elements can be expressed as
\begin{multline}
\Sigma_{\sigma\tau}^{(n)} = N \frac{e^{-W_{\sigma}^{(n)}}-e^{-W_{\tau}^{(n)}}}{W_
{\sigma}^{(n)}-W_{\tau}^{(n)}}\ \times \\\sum_{\substack{i,k\text{ c.c. in
}\sigma\\j,l\text{
    c.c. in }\sigma}} \left(\lambda_i\lambda_j W_{ij}^{(n)} +
\bar{\lambda}_k\bar{\lambda}_l W_{kl}^{(n)}
-\lambda_i\bar{\lambda}_l W_{il}^{(n)} - \bar{\lambda}_k\lambda_j
W_{kj}^{(n)}\right)\delta_{\substack{i,l\text{ c.c. in }\tau\\k,j\text{
      c.c. in }\tau}}
\end{multline}
and
\begin{equation}
W_{\sigma}^{(n)} = -\ln V_\sigma^{(n)}= N\sum_{i,j\text{ c.c. in }\sigma}
\lambda_i\bar{\lambda}_jW_{ij}^{(n)} \ .
\end{equation}
This source of subleading correction contributes when $\sigma$ and
$\tau$ are identical or differ by a single flip. In the latter case
(one flip), we include a factor of $2 \,
\text{Re}\left[\Sigma_{\sigma\tau}^{(n)}\right]$ and the other factor
of $1/N$ comes from the scalar product matrix. In the former case
there are three possibilities. Specifically, we may include either one
or two factors of $\Sigma$ (in either the amplitude or the conjugate
amplitude) in such a way as to undo the effect of a one or two-flip
real emission (see below for the rules for including real emissions),
these contribute to the dark orange and green boxes in the figure. Or
else we may include a factor of $|\Sigma_{\tau\sigma}^{(n)}|^2$ in the
case that the two flips (one from each $\Sigma$) cancel each other out
(green boxes).

At this order we also need to include corrections which are suppressed
by $1/N^2$ and are proportional to ${\mathbf s}\cdot {\mathbf s}$
(light orange boxes) in the virtual evolution operator. This term is
diagonal in colour and
\begin{equation}
{\mathbf V}_n^{\text{NLC},\rho}|\sigma\rangle = \frac{\rho^{(n)}}{N^2}|\sigma\rangle~, ~~~\text{where} ~~~
\rho^{(n)} = N \sum_{i<j}
(\lambda_i-\bar{\lambda}_i)(\lambda_j-\bar{\lambda}_j) W_{ij}^{(n)} \ .
\end{equation}

As discussed above, we must also consider subleading corrections to
real gluon emission. Recall that the $1/N^2$ corrections arise when
$\tau$ and $\sigma$ differ by at most two flips. In this case, we are
to include contributions where the gluon is emitted off either colour
line $c_i$ or $\bar{c}_k$ in the amplitude ($c_i$ and $\bar{c}_k$ are
colour connected in $\sigma\backslash n$), and either colour line
$c_l$ or $\bar{c}_j$ in the conjugate amplitude ($c_l$ and $\bar{c}_j$
are colour connected in $\tau\backslash n$). Evolving towards the hard
process, we are to combine the dipoles $(i,n)$ and $(n,k)$ in
$\sigma$, and $(l,n)$ and $(n,j)$ in~$\tau$, leaving behind dipoles
$(i,k)$ and $(l,j)$. The corresponding factor is \beq \lambda_i
\lambda_l\ \omega_i^{(n)}\left(\omega_l^{(n)}\right)^* +
\bar{\lambda}_k\bar{\lambda}_j\ \omega_k^{(n)}\left(\omega_j^{(n)}\right)^*
-
\lambda_i\bar{\lambda}_j\ \omega_i^{(n)}\left(\omega_j^{(n)}\right)^*
-
\bar{\lambda}_k\lambda_l\ \omega_k^{(n)}\left(\omega_l^{(n)}\right)^* \label{eq:2flip}
\ , \eeq which comes from the first two lines in
Eq.~(\ref{eqn:emittors}). Notice that a potential $1/N$ contribution
arising when $\sigma$ and $\tau$ differ by only one flip vanishes
because of the first bullet point in the list above
Eq.~(\ref{eq:titj}), i.e.\ contributions of the type $\mathbf{t
  [\cdots] t}$ require $\sigma$ and $\tau$ to differ by two
flips\footnote{Recall we are ignoring any off-diagonality due to the
  hard scattering matrix in the way we count flips.}.  If $\sigma$ and
$\tau$ differ by one flip and the gluon connects to itself in $\tau$
but not in $\sigma$, then we should combine the dipoles $(i,n)$ and
$(n,k)$ in $\sigma$ and include a factor of \beq \label{eq:1flip}
-\left(\lambda_i \ \omega_i^{(n)} -
\bar{\lambda}_k\ \omega_k^{(n)}\right)\sum_j
(\lambda_j-\bar{\lambda}_j)\left(\omega_j^{(n)}\right)^* \ , \eeq or
the corresponding conjugate. This corresponds to the third line in
Eq.~(\ref{eqn:emittors}). Again a possible $1/N$ correction arising
when $\sigma$ and $\tau$ are equal vanishes because contributions of
the type $\mathbf{s [\cdots] t}$ require $\sigma$ and $\tau$ to differ
by one flip (see the second bullet point in the list above
Eq.~(\ref{eq:titj})).  Finally, if the gluon is connected to itself in
both $\sigma$ and $\tau$, we include a factor of \beq \sum_{i,j}
(\lambda_i-\bar{\lambda}_i)(\lambda_j-\bar{\lambda}_j) \text{Re}
\left[\omega_{i}^{(n)}\left(\omega_{j}^{(n)}\right)^*\right] \ , \eeq
corresponding to the fourth line in Eq.~(\ref{eqn:emittors}). Armed
with these rules it is possible to go ahead and compute the first
subleading colour contributions to the BMS equation. We leave such a
phenomenological study to future work.

\section{Conclusions}

Accounting systematically for partonic radiation in short-distance
scattering processes is of practical importance and theoretical
interest. Progress in accurately accounting for this physics has been
dominated by coherence-improved parton/dipole shower Monte Carlo
programs \cite{Bellm:2015jjp,Sjostrand:2014zea,Gleisberg:2008ta}
though to date these are all limited to leading $N$, with some subleading
improvements \cite{Platzer:2012np}. Probably
the main challenge in going beyond leading $N$ arises because of the
need to include quantum interference effects, which would seem to
necessitate an amplitude-level approach. This paper represents our
first steps towards the implementation of a general algorithmic approach to
amplitude-level parton evolution, which has also been advocated in
\cite{Nagy:2014mqa}. We anticipate that numerical results using the
technology outlined in this work will be available soon, and we
postpone a detailed discussion of the computational methods to a
follow-up work.

\section*{Acknowledgments}
This work has received funding from the UK Science and Technology Facilities
Council (grant no.\ ST/P000800/1), the European Union's Horizon 2020 research
and innovation programme as part of the Marie Sk\l{}odowska-Curie Innovative
Training Network MCnetITN3 (grant agreement no.\ 722104).  JRF thanks the
Institute for Particle Physics Phenomenology in Durham for the award of an
Associateship. MDA thanks the UK Science and Technology Facilities Council for
the award of a studentship.  SP acknowledges partial support by the COST
action CA16201 PARTICLEFACE, and is grateful for the kind hospitality of ESI
at Vienna, AEC at Bern and MIAPP at Munich, where part of this work has been
addressed. We are indebted to Thomas Becher for valuable discussions. Figures
have been prepared using JaxoDraw \cite{Binosi:2003yf}.

\appendix

\section{The connection with other approaches} \label{ap:others}
In this appendix we show how the colour evolution algorithm defined by
Eq.~\eqref{eq:sigma0}
relates to the previous work of Becher et
al \cite{Becher:2016mmh} and Caron-Huot
\cite{Caron-Huot:2015bja,Caron-Huot:2016tzz}.
\subsection{Becher et al.}
In \cite{Becher:2016mmh}, the hard process $e^+ e^- \to 2$~jets is
considered with the requirement that the total energy emitted outside
of cones centred on the two (back-to-back) jets should satisfy
$2E_\text{out} < \beta Q$ where $Q = \sqrt{s}$. This observable is of
the type described by our Eq.~(\ref{eq:evo3}) and, because there are
no coloured particles in the initial state, the Coulomb terms can be
neglected. Accordingly, in the leading logarithmic approximation they
find (see Section 5.2 of \cite{Becher:2016mmh}).  \bea
\sigma_{\text{LL}}(\delta,\beta) &=& \sigma_0 \, \mathrm{Tr} \,(
      {\boldsymbol{\mathcal{S}}}_2(\{ n_1,n_2\},Q\beta, \delta,
      \mu_h)) \nonumber \\ &=& \sigma_0 \, \sum_{m=2}^{\infty}
      \mathrm{Tr} \,( {\boldsymbol{U}}^S_{2m}(\{n_1,n_2 \},\delta,
      \mu_s, \mu_h) \, \hat{\otimes}
      \,\mathbf{1})~, \label{eq:becher1} \eea where $\delta =
      \tan(\alpha/2)$ ($\alpha$ is the opening angle of the jets),
      $\mu_h = Q$ and $\mu_s = Q\beta$. Formally, the evolution
      operator is given by \beq {\boldsymbol{U}}^S_{lm}(\{n\},\delta,
      \mu_s, \mu_h)= \left(\mathrm{P} \exp \left[ \int_{\mu_s}^{\mu_h}
        \frac{\dd \mu}{\mu} ~ \boldsymbol{\Gamma}^H(\{n\},\delta, \mu)
        \right]\right)_{lm}~, \label{eq:becher2} \eeq where $\{ n \}$
      is the set of $l$ light-like vectors that fix the directions of
      the final-state partons, and \beq \boldsymbol{\Gamma}^H = \left(
\begin{array}{ccccc}
\boldsymbol{\Gamma}^H_{22} & \boldsymbol{\Gamma}^H_{23} &
\boldsymbol{\Gamma}^H_{24} &
\cdots \\
\boldsymbol{\Gamma}^H_{32} & \boldsymbol{\Gamma}^H_{33} &
\boldsymbol{\Gamma}^H_{34} &
\cdots \\
\boldsymbol{\Gamma}^H_{42} & \boldsymbol{\Gamma}^H_{43} &
\boldsymbol{\Gamma}^H_{44} &
\cdots \\
\vdots & \vdots & \vdots  & \ddots
\end{array}
\right)
= \frac{\alpha_s}{4 \pi} \left(
\begin{array}{ccccc}
\boldsymbol{V}_2 & \boldsymbol{R}_2 & 0 & 0 & \cdots \\
0 & \boldsymbol{V}_3 & \boldsymbol{R}_3 & 0 & \cdots \\
0 & 0 & \boldsymbol{V}_4 & \boldsymbol{R}_4 &  \cdots \\
0 & 0 & 0 & \boldsymbol{V}_5 &  \cdots \\
\vdots & \vdots & \vdots & \vdots  & \ddots
\end{array}
\right) + \mathcal{O}(\alpha_s^2) \eeq with \bea \label{eq:bvr}
\boldsymbol{V}_m &=& 2 \sum_{i<j}^{m} ( \mathbf{T}_{i,L} \cdot
\mathbf{T}_{j,L} + \mathbf{T}_{i,R} \cdot \mathbf{T}_{j,R}) \int
\frac{\dd \Omega_k}{4 \pi} \omega_{ij}(\hat{k})~, \nonumber
\\ \boldsymbol{R}_m &=& -4 \sum_{i<j}^m \mathbf{T}_{i,L} \cdot
\mathbf{T}_{j,R} ~ \frac{\dd \Omega_{\hat{q}_{m+1}}}{4 \pi} \,
\omega_{ij}(\hat{q}_{m+1}) \, \Theta_\text{in}(\hat{q}_{m+1})~.  \eea
The $\Theta_\text{in}(\hat{q}_{m+1})$ restricts the emitted gluon with
momentum $q_{m+1}$ to lie inside either the quark or anti-quark jet
(defined by the cones around their directions). There is potential for
confusion here, because this "in" region corresponds to what we called
the "out" region: in both cases we are speaking of the region where
there is no veto on real emissions.  The subscripts "$L$" and "$R$" on
the colour charge operators denote that they sit to the left or right
of the object upon which they operate. Expanding the exponential in
Eq.~(\ref{eq:becher2}) gives rise to exactly the same series as using
Eq.~(\ref{eq:recur}) (see Eq.~(5.17) of \cite{Becher:2016mmh}).

\subsection{Weigert and Caron-Huot}

We can also translate Eq.~(\ref{eq:evo3}) into the notation and
language of \cite{Caron-Huot:2015bja}. The starting point is to
introduce a rotation matrix, $\bold{U}_i$, for each parton in the hard
subprocess and each soft gluon.  Operators $\bold{L}_i$ and
$\bold{R}_i$ are also defined such that
\begin{align}
&\bold{L}_{i}^{a}  \bold{U}_i \equiv \bold{T}_i^a   \bold{U}_i , \qquad
\bold{R}_{i}^{a}  \bold{U}_i \equiv  \bold{U}_i \bold{T}_i^{a\dagger}   \qquad \text{ and}\\
& \bold{L}_{i}^a  \bold{U}_j=    \bold{U}_j \bold{L}_{i}^a \qquad
 \bold{R}_{i}^a  \bold{U}_j = \bold{U}_j \bold{R}_{i}^a, \qquad \text{ for $i\ne j$}.
\end{align}
Their commutation relations are inherited from the colour algebra:
\begin{align}
\begin{gathered}
\left[\bold{R}_j , \bold{L}_k\right]=0,\qquad \left[ \bold{L}_j^{a} ,  \bold{L}_k^{b}
\right] = \delta_{jk} if^{abc}  \bold{L}_j^c,\qquad
 \left[\bold{R}_j^a,  \bold{R}_k^b\right]=- \delta_{jk} if^{abc} \bold{R}_j^{c}.
\end{gathered}\label{eq:LRdef}
\end{align}
Now one defines the one-loop kernel
\begin{eqnarray} \label{eq:ckern}
\bold{K}^{(1)}_{ij}  &=& \frac{\alpha_s \, \mu^{2\ep} }{  \pi (2\pi)^{-2\epsilon}}
\frac{\dd \Omega_{q}^{3-2\epsilon}}{4\pi }
\frac{\left(\omega_{ij}(\hat{q})\right)^{\eta}}{2}   \Bigg[
\omega_{ij}(\hat{q})  \theta_{ij}(q) \nonumber \\ && \times \left(- \, \bold{R}_{i}^a  \bold{U}_q^{ab}  \bold{L}_{j}^b
- \, \bold{R}_{j}^a \bold{U}_q^{ab} \bold{L}_{i}^b
+\bold{R}_{i}^a \bold{R}_{j}^a
+\bold{L}_{i}^a\bold{L}_{j}^a \right)
\nonumber \\
 & &
-i \pi \widetilde{\delta}_{ij}{\frac{\Omega^{2-2\epsilon}}{\Omega^{3-2\epsilon} 2\pi}}
 \left(   \bold{R}_{i}^a \bold{R}_{j}^a-\bold{L}_{i}^a
\bold{L}_{j}^a\right)  \Bigg] ~,
\end{eqnarray}
where $\eta\to 0$ and $\theta_{ij}\to 1$ gives rise to ordering in
energy. Ordering in dipole transverse momentum is obtained with
$\eta\to -\epsilon$ and $\theta_{ij}\to \Theta(p_i\cdot(p_j- q)>0)
\Theta(p_j\cdot(p_i- q)>0)$.  The corresponding equation for a general
observable (i.e.\  Eq.~\eqref{eq:eord} for energy ordering and
Eq.\eqref{eq:kord} for dipole transverse momentum ordering) is \beq
\dd \sigma = \left. \left[{\text{P}}\,\, \text{exp} \left( \sum_{i\ne
    j} \frac{\dd \lambda}{\lambda^{1+2\ep}}\bold{K}^{(1)}_{ij}\right)
  \bra{\mathcal{M}} \bold{U}_1 \cdots \bold{U}_N \ket{\mathcal{M}}
  \right]~ \right|_{\bold{U}_1\cdots \bold{U}_{N+m}\to \mathbb{I}_{N+m}
 }\label{eq:1Lcaron}~, \eeq which is fully differential. Here $N$
is the number of partons in the hard subprocess, the path ordering acts
over $\lambda$ and the colour matrices $U_i$ should be independent of
this parameter. In proving the equivalence of Eq. \eqref{eq:1Lcaron}
with Eq.~({\ref{eq:eord}}) and Eq.~({\ref{eq:kord}}) it is useful to
note that
\begin{eqnarray}
&\text{P}& \frac{\dd \lambda_n}{\lambda_n^{1+2\epsilon}} \bold{K}^{(1)}_{ij}  =
\frac{1}{2}    \Bigg[  \dd \Pi_{q_n} \left( \, \bold{R}_{i}^a  \bold{U}_{q_n}^{ab}  \bold{L}_{j}^b
+ \, \bold{R}_{j}^a \bold{U}_{q_n}^{ab} \bold{L}_{i}^b
-\bold{R}_{i}^a \bold{R}_{j}^a
-\bold{L}_{i}^a\bold{L}_{j}^a \right) \omega_{ij}  (\hat{q}_n)\, \theta_{ij}(q_n) \;
\nonumber \\
&-& i \pi
\frac{\alpha_s \, \mu^{2\ep} }{  \pi (2\pi)^{-2\epsilon}}
 \widetilde{\delta}_{ij}{\frac{\Omega^{2-2\epsilon}}{ 2\pi}}
 \frac{\dd \Omega_{q_{n}}^{3-2\epsilon}}{\Omega_{q_{n}}^{3-2\epsilon}}
 \frac{\dd \lambda_{n}}{\lambda^{1+2\epsilon}_{n}}
 \left(   \bold{R}_{i}^a \bold{R}_{j}^a-\bold{L}_{i}^a
\bold{L}_{j}^a\right) \Bigg]  \Theta(\lambda_{n-1}< \lambda_{n}<\lambda_{n+1} )  ~,
\end{eqnarray}
here $\lambda_{n}=E_n$ for energy ordering and
$\lambda_{n}=q^{(ij)}_n$ for dipole ordering (see below). Up to the
Coulomb gluon term, this is equal to the lowest order resummation
contained in equations (2.7) and (2.14) of
\cite{Caron-Huot:2015bja}. This is also very closely related to the
work of Weigert \cite{Weigert:2003mm}.

Another way to write Eq.~\eqref{eq:1Lcaron} is via a generating
functional:
\beq
Z = \text{P} \exp \left( \sum_{i\ne j}
   \frac{\dd \lambda}{\lambda^{1+2\ep}}\bold{K}^{(1)}_{ij}\right)~ Z_0~,
\eeq
where $Z_0 = \bra{\mathcal{M}}
\bold{U}_1 \cdots \bold{U}_N   \ket{\mathcal{M}}$ and
\beq
 \sigma_m = \int u_m \, \left. \frac{\delta Z}{\delta U_1 \cdots \delta U_{n+m}}
\right|_{\{
U_i \} = 0}~.
\eeq

The case $\lambda_{n}=q^{(ij)}_n$ is reminiscent of (but not the same
as) the dipole transverse momentum ordering we
discussed in Section \ref{sec:ordering}
(see Eq.~(\ref{eq:dipdef1})). Indeed, Eq.~(\ref{eq:1Lcaron}) can be
re-written as a recurrence relation:
\begin{eqnarray}
\bold{A}^{\{(i_1, j_1),\dots ,(i_n, j_n)\}}_{n}(\mu) \equiv  & &   \\
& & \hspace{-4cm} ~~~~~~~~~~~~~~\bold{V}_{\mu  q_n^{(i_{n}j_{n})}}\!
\left[  \theta_{i_nj_n}(q_n) \bold{T}_{i_n}
\bold{A}^{\{(i_1, j_1),\dots ,(i_{n-1}, j_{n-1})\}}_{n-1}\left(q_n^{(i_{n}j_{n})}\right)
  \bold{T}_{j_n}  \omega_{i_n j_n}(\hat{q}_n)\right]
  \bold{V}^\dagger_{\mu q_n^{(i_{n}j_{n})}}~, \nonumber \label{eq:ch}
\end{eqnarray}
where $\mathbf{A}_0(\mu) = \mathbf{V}_{\mu,Q} \, \mathbf{H} \, \mathbf{V}^\dag_{\mu,Q}$
 and $\mathbf{V}$ is defined analogously to Eq.~\eqref{eq:simple}.
Formally, each of the gluons should have an
energy $E < Q$ and this is imposed
via $\theta_{i_nj_n}(q_n)$. In \cite{Angeles-Martinez:2016dph},
direct calculation led to $ \theta_{ij}(q)  = \Theta(p_{i}\cdot (p_{j}-q)>0)
\Theta(p_{j}\cdot (p_i-q)>0)$ and we introduce
it here to cut-off arbitrarily high momentum
modes. We cannot avoid the long chain of
indices because the observable is
obtained by integrating over the
multi-gluon
phase space subject to $\Theta\left(q_1^{(i_1,j_1)}>q_2^{(i_2,j_2)}> \cdots\right)$, i.e.
\begin{eqnarray}
	\sigma &=&  \sum_n \, \Bigg[ \left( \prod_{i=1}^n \int \dd\Pi_i \right) \,
\sum_{i_1,j_1}^N \sum_{i_2,j_2}^{N+1} \cdots \sum_{i_n,j_n}^{N+n-1}  ~    \,
\Theta\left(q_1^{(i_1j_1)}>q_2^{(i_2j_2)}> \cdots\right) ~ u_n(k_1,k_2,\cdots,k_n)
\nonumber  \\
& \times &  \text{Tr} \bold{A}^{\{(i_1, j_1),\dots ,(i_n, j_n)\}}_{n}(\mu)
\Bigg] ~, \label{eq:kord}
\end{eqnarray}
where $i_m,~j_m \leq N+m-1$ and $N$ is the number of hard partons.  As pointed
out in \cite{Caron-Huot:2015bja,Caron-Huot:2016tzz}, choosing the dipole
transverse momentum to order the emissions is, ultimately, a renormalization
scheme choice in the effective theory, albeit one that has the virtue of
making Lorentz invariance manifest.  The dipole transverse momenta of
successive real emissions are ordered and this set of ordered momenta acts to
limit the virtual gluon loop integrals in Eq.~(\ref{eq:ch}).

We note that dipole ordering avoids all collinear poles except for those
associated with the very last emission. This is a very attractive feature. The
proof proceeds along the following lines: for a given dipole chain, poles come
from $(p_i \cdot q) \, (p_j \cdot q) = 0$. But this quantity is proportional
to the ordering variable, so the only possibility of it equalling zero is the
case of the final emission (when $\mu = 0$).

\section{On the cancellation of infrared divergences below the inclusivity
scale}
\label{ap:BN}

The aim here is to show that, for observables fully inclusive for $E < \rho$, we
can simply impose $E > \rho$ in the algorithm. This fact follows because
\begin{align}
\sigma = \text{Tr} \, \mathbf{H} + \sum_{n=0}^{\infty}
& \int  \left( \prod_{m=1}^{n+1} \dd \Pi_m\right)  \,
\text{Tr} \, \bold{A}_{n+1}(E_{n+1}) \, \left ( u_{n+1}(q_1,\dots,q_{n+1})-u_{n}(q_1,\dots,q_{n})\right)
,\label{eq:BN}
\end{align}
hence if $u_{n+1}= u_{n}$ for $E_{n+1}<\rho$ then one can set $\rho$ as the lower
bound on the energy integrals for both real emissions and virtual exchanges.

To prove Eq.~(\ref{eq:BN}) we make use of the identity
\begin{align}
 \bold{V}^{\dagger}_{a,b}\bold{V}_{a,b}-1 = \frac{\alpha_s}{\pi} \int_a^b
 \frac{\dd
 E}{E} \frac{\dd \Omega_q}{4\pi}
\bold{V}_{E,b}^\dagger  \bold{D}^2(\hat{q}) \bold{V}_{E,b}~,
\label{eq:1stcurrentidentity}
\end{align}
where we used the shorthand $\bold{D}^2(\hat{q})\equiv
\bold{D}^{\mu}(q)\, \bold{D}_{\mu}(q)$
and the Sudakov operator is given in Eq.~\eqref{eq:simple}.
Using Eq.~\eqref{eq:1stcurrentidentity}, we can rewrite the
contribution to the observable from $n$ real emissions as
\begin{align}
\sigma_n
&= \int \left (\prod_{m=1}^{n}\dd \Pi_m\right)  \,
\text{Tr} \,
\mathbf{V}_{E_n,0} \mathbf{A}_{n}(E_n) \mathbf{V}_{E_n,0}^\dag\, \, 
\label{eq:bn2}  \\
&=\int \left (\prod_{m=1}^{n}\dd \Pi_m\right)
\,\text{Tr} \Big(
  \mathbf{A}_{n}(E_n)     -\int \dd \Pi_{n+1} \,
  \mathbf{A}_{n+1}(E_{n+1})
   \Big),\nonumber
\end{align}
where it should be understood that $E_0 =Q$, i.e.
$\mathbf{A}_0(E_0)=\mathbf{H}$.
Eq.~\eqref{eq:BN} trivially follows from this
expression by grouping terms that depend on the same trace.

\section{Non-global logarithms at fixed order in $\alpha_s$} \label{app:fo}
\subsection{Calculation of $\Sigma_1$ at order $\alpha_s^2$}

In this appendix, we compute the fixed-order expansion of the non-global
logarithmic contributions to the hemisphere mass. Apart from checking the
correctness of the algorithm, this allows us to confirm that the expansion
proposed in Section \ref{sec:ngo} is indeed free from infra-red divergences at
each order, i.e.\ the $\Sigma_n$ are separately finite. We will be very
explicit in the hope that it will be useful to see how a calculation proceeds
in detail.

First we compute the hemisphere jet mass to fixed order, as in Dasgupta-Salam
\cite{Dasgupta:2001sh}. As they do, we start by computing the lowest order
non-global correction to the cumulative event shape where the jet mass is
required to be less than $\rho$.  We can do this by using the algorithm with
$\mu = \rho$ and taking the "out" region to be the region of phase space that
does not contribute to the hemisphere jet mass, i.e.\ it is the wrong-side
hemisphere. The "in" region is the complement of this. Note it is only to
leading $\ln (Q/\rho)$ accuracy that the observable is fully inclusive over
gluon emissions with $E < \rho$. In which case we may write (see
Eq.~(\ref{eq:sngl}) with $\mathbf{\wb{V}} \to \mathbf{V}_{\text{in}}$ and
$-\tfrac{1}{2} \mathbf{D}^2 \to \boldsymbol{\gamma}$): \bea \Sigma_{1}(\rho)
&=& \frac{1}{\sigma} \int_0^\rho \frac{\dd \sigma}{\dd \rho} \dd \rho =
\frac{1}{N}\int_{\text{out}} \dd \Pi_1~ \left[ \mathrm{Tr}(
  {\mathbf{V}}_{\rho,E_1}^{\text{in}} \mathbf{D}_1^\mu
  {\mathbf{V}}_{E_1,Q}^\text{in}
  {\mathbf{V}}^{\text{in}\dag}_{E_1,Q}\mathbf{D}_{1\mu}^\dag
  {\mathbf{V}}^{\text{in}\dag}_{\rho,E_1}) + \nonumber \right. \\ & & \left.
  \mathrm{Tr}( {\mathbf{V}}_{\rho,E_1}^\text{in} \boldsymbol{\gamma}_1
         {\mathbf{V}}_{E_1,Q}^\text{in} {\mathbf{V}}^{\text{in}\dag}_{\rho,Q})
         + \mathrm{Tr}( {\mathbf{V}}_{\rho,Q}^\text{in}
         {\mathbf{V}}^{\text{in}\dag}_{E_1,Q}\boldsymbol{\gamma}_1^\dag
         {\mathbf{V}}^{\text{in}\dag}_{\rho,E_1}) \right] ~.  \eea We have set
the Born matrix element equal to the identity (since we are considering a
two-jet $e^+e^-$ event shape) and the factor of $1/N$ removes the colour
factor for the lowest order cross section. To lowest order, \bea
{\mathbf{V}}^\text{in}_{a,b} \approx 1 -\frac{\alpha_s}{\pi} \int_{a}^{b}
\frac{\dd E}{E} \sum_{i<j} (-\bold{T}_i \cdot \bold{T}_j) \int_{\text{in}}
\frac{\dd \Omega}{4\pi} \omega_{ij}~.  \eea Expanding out gives (note the
lower case notation, $\mathbf{t}_a$, which (only in this appendix) indicates
that these operators act on 3-parton objects): \bea \Sigma_{1}(\rho) &=&
\left(\frac{\alpha_s}{\pi}\right)^2 \frac{1}{N} \int_{\text{in}} \frac{\dd
  \Omega}{4\pi} \int_\rho^Q\frac{\dd E_1}{E_1} \int_{\text{out}} \frac{\dd
  \Omega_1}{4\pi} \Big[ \nonumber \\ & & - 4\omega_{ab}(q_1)\int_{\rho}^{E_1}
  \frac{\dd E}{E} ~ \mathrm{Tr}\left( ( \mathbf{t}_a \cdot \mathbf{t}_{q_1} \,
  \omega_{aq_1}(k) + \mathbf{t}_b \cdot \mathbf{t}_{q_1} \, \omega_{bq_1}(k)
  +\mathbf{t}_a \cdot \mathbf{t}_b \, \omega_{ab}(k) ) \;
  \mathbf{T}_a\cdot\mathbf{T}_b \right) \nonumber \\ & & -
  4\omega_{ab}(q_1)\int_{E_1}^Q \frac{\dd E}{E} \; \omega_{ab}(k) \;
  \mathrm{Tr}( \mathbf{T}_a\cdot \mathbf{T}_b \;\mathbf{T}_a \cdot
  \mathbf{T}_b ) \nonumber \\ & & + 2\omega_{ab}(q_1)\int_\rho^{E_1} \frac{\dd
    E}{E} \; \omega_{ab}(k) \; \mathrm{Tr}( \mathbf{T}_a\cdot \mathbf{T}_b \;
  \mathbf{T}_a\cdot \mathbf{T}_b ) \nonumber \\ & & +
  2\omega_{ab}(q_1)\int_{E_1}^Q \frac{\dd E}{E} \; \omega_{ab}(k) \;
  \mathrm{Tr}( \mathbf{T}_a\cdot \mathbf{T}_b \; \mathbf{T}_a\cdot
  \mathbf{T}_b ) \nonumber \\ & & + 2\omega_{ab}(q_1)\int_\rho^{Q} \frac{\dd
    E}{E} \; \omega_{ab}(k) \; \mathrm{Tr}( \mathbf{T}_a\cdot \mathbf{T}_b \;
  \mathbf{T}_a\cdot \mathbf{T}_b ) \Big] \nonumber \\ \eea which reduces
nicely to \bea \Sigma_{1}(\rho) &=& -\left(\frac{\alpha_s}{\pi} \right)^2
\frac{4}{N} \left( \int_\rho^Q\frac{\dd E_1}{E_1} \int_{\text{out}} \frac{\dd
  \Omega_1}{4\pi}\right) \left(\int_\rho^{E_1} \frac{\dd E_k}{E_k}
\int_{\text{in}} \frac{\dd \Omega_k}{4\pi}\right) \\ & \times &
\omega_{ab}(q_1)~ \mathrm{Tr}\left[ ( \mathbf{t}_a \cdot \mathbf{t}_{q_1} \,
  \omega_{aq_1}(k) + \mathbf{t}_b \cdot \mathbf{t}_{q_1} \, \omega_{bq_1}(k)
  +(\mathbf{t}_a \cdot \mathbf{t}_b - \mathbf{T}_a \cdot \mathbf{T}_b) \,
  \omega_{ab}(k))\mathbf{T}_a\cdot\mathbf{T}_b \right]\nonumber ~.  \eea To
compare to \cite{Dasgupta:2001sh} we write \bea p_a &=& \frac{Q}{2}(1,0,0,1)
\\ p_b &=& \frac{Q}{2}(1,0,0,-1) \nonumber \\ q_1 &=& x_1 \frac{Q}{2}(1,0,\sin
\theta_1, \cos\theta_1) \nonumber \\ k &=& x_2 \frac{Q}{2}(1,\sin\theta_2 \sin
\phi,\sin\theta_2 \cos \phi,\cos\theta_2) \nonumber \eea and \bea
\omega_{aq_1}(k) &=& \frac{(1 - \cos \theta_1)}{(1-\cos \theta_2) \; (1-\sin
  \theta_1 \sin \theta_2 \cos \phi - \cos \theta_1 \cos\theta_2)} \nonumber
\\ \omega_{bq_1}(k) &=& \frac{(1 + \cos \theta_1)}{(1+\cos \theta_2) \;
  (1-\sin \theta_1 \sin \theta_2 \cos \phi - \cos \theta_1 \cos\theta_2)}
\nonumber \\ \omega_{ab}(q_1) &=& \frac{2}{\sin^2 \theta_1} \nonumber
\\ \omega_{ab}(k) &=& \frac{2}{\sin^2 \theta_2}~.  \nonumber \\ \eea We can do
the azimuthal integral using
\begin{align}
\int_0^{2 \pi} \frac{\dd \phi}{2 \pi} \, \frac{1}{(1-\sin \theta_1 \sin
	\theta_2 \cos \phi - \cos \theta_1 \cos\theta_2)} = \frac{1}{|\cos \theta_1
	-
	\cos \theta_2|}
\end{align}
then
\bea
\Sigma_{1}(\rho) &=& -\left(\frac{\alpha_s}{\pi} \right)^2 \frac{2}{N}
\left( \int^1_{\rho/Q} \frac{\dd x_1}{x_1} \int_{-1}^0
\frac{\dd(\cos\theta_1)}{\sin^2 \theta_1} \ \right) \times \left(
\int_{\rho/Q}^{x_1} \frac{\dd x_2}{x_2} \int_0^1
\frac{\dd(\cos\theta_2)}{\sin^2\theta_2} \right) \nonumber \\ & &
\times  \Bigg[ \mathrm{Tr}(  \mathbf{t}_a \cdot \mathbf{t}_{q_1} \;
\mathbf{T}_a\cdot\mathbf{T}_b) \; \frac{(1-\cos \theta_1)(1+\cos
	\theta_2)}{\cos \theta_2 - \cos \theta_1}    \nonumber \\
& & ~~~+\mathrm{Tr}(\mathbf{t}_b \cdot \mathbf{t}_{q_1} \;
\mathbf{T}_a\cdot\mathbf{T}_b) \; \frac{(1-\cos \theta_2)(1+\cos
	\theta_1)}{\cos \theta_2 - \cos \theta_1}  \;   \nonumber \\
& & ~~~+ 2\, \mathrm{Tr}((\mathbf{t}_a \cdot \mathbf{t}_b - \mathbf{T}_a \cdot
\mathbf{T}_b)\; \mathbf{T}_a\cdot\mathbf{T}_b)
\Bigg]~.
\eea
Now do the colour traces, i.e.
\bea
\text{Tr}(\mathbf{t}_a \cdot \mathbf{t}_q \; \mathbf{T}_a\cdot\mathbf{T}_b) &=&
\text{Tr}(\mathbf{t}_b \cdot \mathbf{t}_q \; \mathbf{T}_a\cdot\mathbf{T}_b) =
NC_F^2 + \frac{C_F}{2}~,
\nonumber \\
\text{Tr}(\mathbf{t}_a \cdot \mathbf{t}_b \; \mathbf{T}_a\cdot\mathbf{T}_b) &=&
-\frac{C_F}{2}~, \nonumber \\
\text{Tr}(\mathbf{T}_a \cdot \mathbf{T}_b \; \mathbf{T}_a\cdot\mathbf{T}_b) &=&
NC_F^2~.
\eea
So that
\bea
\Sigma_{1}(\rho) &=& -N C_F \left(\frac{\alpha_s}{\pi} \right)^2
\left( \int_{\rho/Q}^1 \frac{\dd x_1}{x_1} \int_{-1}^0
\frac{\dd(\cos\theta_1)}{\sin^2 \theta_1}   \;\right) \left(
\int_{\rho/Q}^{x_1} \frac{\dd x_2}{x_2} \int_0^1
\frac{\dd(\cos\theta_2)}{\sin^2\theta_2}   \right) \\
& & \times  \; \frac{1}{\cos \theta_2 - \cos \theta_1} \Bigg[ (1-\cos
\theta_1)(1+\cos \theta_2)   + (1-\cos \theta_2)(1+\cos \theta_1) - 2(\cos
\theta_2 - \cos \theta_1)
\Bigg]~~ \nonumber \\
&\approx & -\left(\frac{\alpha_s}{\pi} \right)^2 \ln^2(Q/\rho)
\left( \int_{-1}^0 \frac{\dd(\cos\theta_1)}{\sin^2 \theta_1} \; \right)
\left( \int_0^1 \frac{\dd(\cos\theta_2)}{\sin^2\theta_2} \; \right)
\frac{NC_F}{\cos \theta_2 - \cos \theta_1} \nonumber \\ & &
\times  (1-\cos \theta_2)(1+\cos \theta_1) ~. \nonumber
\eea
Using
\bea
\int_{-1}^0 \dd x \;  \int_0^1 \dd y \; \frac{1}{(y-x)(1-x)(1+y)} &=&
\frac{\zeta(2)}{2}
\eea
gives
\bea
\Sigma_{1}(\rho)
&\approx & -\frac{NC_F}{2} \zeta(2) \left(\frac{\alpha_s}{\pi} \right)^2 \ln^2 (Q/\rho)
~,
\eea
which is equal to the result in \cite{Dasgupta:2001sh}.

\subsection{Calculation of ${\Sigma_1}$ and ${\Sigma_2}$ at order ${\alpha_s^3}$}
The same methodology as in the previous subsection can be used to
compute $\Sigma_2$ at order $\alpha_s^3$, and $\Sigma_1$ at
the same order. The sum $\Sigma_1 + \Sigma_2$ then gives the
non-global contribution at order $\alpha_s^3$. The result for
$\Sigma_2$ is
\begin{align}
\Sigma_2 (\rho) &= \frac{1}{N} \int_{\text{out}} \frac{\alpha_s}{\pi} \int_{\rho}^Q \frac{\dd E_1}{E_1} \frac{\dd \Omega_1}{4 \pi} \int_{\text{out}} \frac{\alpha_s}{\pi} \int_{\rho}^{E_1} \frac{\dd E_2}{E_2} \frac{\dd \Omega_2}{4 \pi} \frac{\alpha_s}{\pi} \int_{\text{in}} \frac{\dd \Omega_k}{4 \pi} \int_{\rho}^{E_2} \frac{\dd E_k}{E_k} \Big[ \nonumber \\
&- 4 C_A^2 C_F^2 \left(\omega_{ab}(q_1) \omega_{ab}(q_2) \omega_{a q_1}(k) + \omega_{ab}(q_1) \omega_{ab}(q_2) \omega_{b q_1}(k) \right) \nonumber \\
&-  2 C_A C_F \left( \omega_{ab}(q_1) \omega_{ab}(q_2) \omega_{a q_2}(k) + \omega_{ab}(q_1) \omega_{ab}(q_2) \omega_{b q_2}(k) \right)\nonumber \\
&+ 2 C_A^3 C_F \left( \omega_{ab}(q_1) \omega_{ab}(q_2) \omega_{ab}(k) - \omega_{ab}(q_1) \omega_{a q_1}(q_2) \omega_{a q_1}(k) \right. \nonumber \\
&+ \left. \omega_{ab}(q_1) \omega_{a q_1}(q_2) \omega_{a q_2}(k) + \omega_{ab}(q_1) \omega_{a q_1}(q_2) \omega_{q_1 q_2}(k) - \omega_{ab}(q_1) \omega_{b q_1}(q_2) \omega_{b q_1}(k) \right. \nonumber \\
& \left. + \omega_{ab}(q_1) \omega_{b q_1}(q_2) \omega_{b q_2}(k) + \omega_{ab}(q_1) \omega_{b q_1}(q_2) \omega_{q_1 q_2}(k) \right) \Big] \nonumber \\
&= -\frac{1}{N} \int_{\text{out}} \frac{\alpha_s}{\pi} \int_{\rho}^Q \frac{\dd E_1}{E_1} \frac{\dd \eta_1 \dd \phi_1}{4 \pi} \int_{\text{out}} \frac{\alpha_s}{\pi} \int_{\rho}^{E_1} \frac{\dd E_2}{E_2} \frac{\dd \eta_2 \dd \phi_2}{4 \pi} \frac{\alpha_s}{\pi} \int_{\text{in}} \frac{\dd \eta_k \dd \phi_k}{4 \pi} \int_{\rho}^{E_2} \frac{\dd E_k}{E_k} \Big[ \nonumber \\
& 2 C_A^3 C_F \omega_{ab}(q_1) ( A_{a1}^{q_2 k} + A_{q_1 b}^{q_2 k} - A_{ab}^{q_2 k} ) + \frac{4 C_A^2 C_F^2}{2} \omega_{ab}(q_1) \omega_{ab}(q_2) ( A_{ab}^{q_2 k} - A_{ab}^{q_1 k} ) \Big] \nonumber \\
&= - \left(\frac{\alpha_s}{\pi}\right)^3 \frac{\ln(Q/\rho)^3}{3!} C_A^2 C_F \zeta_3~.
\label{eqn:sigma2}
\end{align}
In order to facilitate comparison with the work of Delenda \& Khelifa-Kerfa \cite{Khelifa-Kerfa:2015mma}, we have used the notation
\begin{equation}
A_{ab}^{ij} = \omega_{ab}(q_i) ( \omega_{a q_i}(q_j) + \omega_{q_i b}(q_j) - \omega_{ab}(q_j) )~,
\end{equation}
see Eq.~(2.2b) of \cite{Khelifa-Kerfa:2015mma}. In addition, $\Sigma_1$ at order $\alpha_s^3$ is
\begin{align}
\Sigma_1^{\alpha_s^3} (\rho) &= \frac{1}{N} \int_{\text{out}} \frac{\alpha_s}{\pi} \int_{\rho}^Q \frac{\dd E_1}{E_1} \frac{\dd \Omega_1}{4 \pi} \int_{\text{in}} \frac{\alpha_s}{\pi} \int_{\rho}^{E_1} \frac{\dd E_2}{E_2} \frac{\dd \Omega_2}{4 \pi} \frac{\alpha_s}{\pi} \int_{\text{in}} \frac{\dd \Omega_k}{4 \pi} \int_{\rho}^{E_2} \frac{\dd E_k}{E_k} \Big[ \nonumber \\
& C_A C_F \left( \omega_{ab}\left(q_1\right) \omega_{ab}\left(q_2\right) \left( \omega_{a q_1}\left(k\right) + \omega_{b q_1}\left(k\right) - \omega_{ab}\left(k\right) \right) \right. \nonumber \\
&+ \left.  \omega_{ab}\left(q_1\right) \omega_{ab}\left(k\right) \left( \omega_{a q_1}\left(q_2\right) + \omega_{b q_1}\left(q_2\right) - \omega_{ab}\left(q_2\right) \right) \right) \nonumber \\
&+ C_A^3 C_F \,  \omega_{ab} \left(q_1\right) \left( \omega_{ab} \left(q_2\right) \omega_{ab} \left(k\right) - \omega_{a1} \left(q_2\right) \omega_{a q_1} \left(k\right) \right. \nonumber \\
&- \left. \omega_{b q_1} \left(q_2\right) \omega_{b q_1} \left(k\right) - \omega_{b q_1} \left(q_2\right) \omega_{a q_1} \left(k\right) - \omega_{a q_1} \left(q_2\right) \omega_{b q_1} \left(k\right) \right) \Big] \nonumber \\
&= -\frac{1}{N} \int_{\text{out}} \frac{\alpha_s}{\pi} \int_{\rho}^Q \frac{\dd E_1}{E_1} \frac{\dd \eta_1 \dd \phi_1}{4 \pi} \int_{\text{in}} \frac{\alpha_s}{\pi} \int_{\rho}^{E_1} \frac{\dd E_2}{E_2} \frac{\dd \eta_2 \dd \phi_2}{4 \pi} \frac{\alpha_s}{\pi} \int_{\text{in}} \frac{\dd \eta_k \dd \phi_k}{4 \pi} \int_{\rho}^{E_2} \frac{\dd E_k}{E_k} \Big[ \nonumber \\
& -C_A^3 C_F A_{ab}^{q_1 q_2} \bar{A}_{ab}^{q_1 k} - 2 C_A^2 C_F^2 ( \omega_{ab}^3 A_{ab}^{q_1 q_2} + \omega_{ab}^2 A_{ab}^{q_1 k} ) \Big] \nonumber \\
&= 2 \left(\frac{\alpha_s}{\pi}\right)^3 \frac{\ln(Q/\rho)^3}{3!} C_A^2 C_F \, \zeta(3)~,
\label{eqn:sigma1alphas3}
\end{align}
where $\bar{A}_{ab}^{q_i q_j} = A_{ab}^{q_i q_j} / \omega_{ab}(q_i)$ (see Eq.~(3.8) and Eq.~(3.11) in \cite{Khelifa-Kerfa:2015mma}). This is in agreement with the result in \cite{Khelifa-Kerfa:2015mma,Delenda:2015tbo}, which is written as
\begin{align}{}
\Sigma_1^{\alpha_s^3} &= - \int_{x_1 > x_2 > x_3} ~\mathrm{d}  \Pi_{123} ~\theta_1^{\text{out}} \theta_2^{\text{in}} \theta_3^{\text{in}} \,\overline{\mathcal{W}}^{RVR}_{123} + \Sigma_3^B \nonumber \\
\Sigma_2 &= - \int_{x_1 > x_2 > x_3} ~\mathrm{d} \Pi_{123}~ \theta_1^{\text{out}} \theta_2^{\text{out}} \theta_3^{\text{in}} \, \left(\overline{\mathcal{W}}^{RRR}_{123} + \overline{\mathcal{W}}^{RVR}_{123} \right)~,
\end{align}
where $\theta_i^{\text{in}}= \theta(\eta_i),~ \theta_i^{\text{out}}= \theta(-\eta_i)$ and  explicit expressions for $\overline{\mathcal{W}}^{RVR}_{123}$ and $\overline{\mathcal{W}}^{RRR}_{123}$ are presented in \cite{Delenda:2015tbo}.

\section{Working in a non-orthogonal colour basis} \label{sec:dual}
Generally we wish to compute \beq \langle H | \mathbf{V}^\dag \mathbf{V} | H
\rangle = \text{Tr}(\mathbf{O}) \eeq where $\mathbf{O} = \mathbf{V} | H
\rangle \langle H| \mathbf{V}^\dag$, and $|H\rangle$ represents the hard
scattering process while $\mathbf{V}$ accounts for the subsequent evolution
(real and virtual). Since the basis is non-orthonormal it is useful to
introduce dual basis vectors, $| \alpha ]$ defined so that \bea \sum_\alpha
|\alpha \rangle [ \alpha | &=& \sum_\alpha |\alpha] \langle \alpha | =
\mathbf{1} ~~~~~\text{and} \nonumber \\ \langle \alpha | \beta ] &=& [ \alpha
  | \beta \rangle = \delta_{\alpha \beta} ~.  \eea We can now write \bea
  \text{Tr}(\mathbf{O}) &=& \text{Tr}( \, [ \sigma | \mathbf{O} | \tau ] \,
  \langle \tau | \sigma \rangle \, ) ~~~~~ \text{and} \\ \mathbf{O} &=&
  \sum_{\sigma, \tau} [ \sigma | \mathbf{O} | \tau ] \, | \sigma \rangle
  \langle \tau |~.  \eea Our interest is to compute the matrix elements $[
    \sigma | \mathbf{O} | \tau ] $ for a specified pair of external states,
  $\sigma$ and $\tau$. This we do by evolving inwards from the external
  states, stripping off soft-gluon operators as we head towards the hard
  scattering (which lies at the heart of $\mathbf{O}$). The key result in
  allowing us to accomplish this is the fact that we can write $\mathbf{O} =
  \mathbf{L} \mathbf{O}' \mathbf{R}$ where $\mathbf{L}$ and $\mathbf{R}$ are
  colour reconnectors, which means \beq \mathbf{R} | \alpha \rangle =
  C_R^\alpha \, | \beta \rangle~~~~~~\text{where} ~~~~ C_R^\alpha = [ \beta |
    \mathbf{R} | \alpha \rangle~.  \eeq Note that there is no sum over $\beta$
    on the right-hand side, i.e.\ reconnectors constitute a unique map from
    one basis vector into another. A similar relation holds for
    $\mathbf{L}$. To make the equations slightly simpler, we will put
    $\mathbf{L} = \mathbf{1}$ in what follows. We want to calculate \bea [
      \sigma | \mathbf{O}' \mathbf{R} | \tau ] &=& \sum_\alpha [ \sigma |
      \mathbf{O}' | \alpha] \langle \alpha | \mathbf{R} | \tau ] \nonumber
  \\ &=& \sum_\alpha [ \sigma | \mathbf{O}' | \alpha] ( [ \tau |
    \mathbf{R}^\dag | \alpha \rangle )^* \\ &=& \sum_\alpha [ \sigma |
      \mathbf{O}' | \alpha] ( C_R^{\alpha*} \, [ \tau | \beta \rangle )^*
      \\ &=& [ \sigma | \mathbf{O}' | \alpha] \, C_R^\alpha~, \eea where in
      the final line the state $\alpha$ satisfies $\mathbf{R} | \alpha \rangle
      = C_R^\alpha | \tau \rangle $. This state $\alpha$ is unique since the
      state $\tau$ is fixed. In this way, we see that it is possible to
      recursively strip off evolution operators leaving behind $c$-number
      factors and reduced matrix elements in the dual basis.

In the particular case of a real gluon emission (see Eq.~(\ref{eqn:emittors}))
we must evaluate \bea [\sigma|{\mathbf T}_i \, {\mathbf A} \, {\mathbf
    T}_j|\tau ] & = & \sum_{\alpha, \beta} [\sigma|{\mathbf T}_i |\alpha
  \rangle \, \langle \beta | {\mathbf T}_j |\tau ] \, [ \alpha | {\mathbf A} |
  \beta ] \, , \eea where \bea [\sigma|{\mathbf T}_i |\alpha \rangle \,
  \langle \beta | {\mathbf T}_j |\tau ] &=&\delta_{\alpha, \sigma\backslash n}
\left(\lambda_i\delta_{c_i\sigma^{-1}(\bar{c}_n)}-\bar{\lambda}_i\delta_{\bar{c}_i\sigma(c_n)}
-\frac{1}{N}(\lambda_i-\bar{\lambda}_i)\delta_{c_n\sigma^{-1}(\bar{c}_n)}\right)\\ &
& \times\left(\lambda_j\delta_{c_j\tau^{-1}(\bar{c}_n)}-
\bar{\lambda}_j\delta_{\bar{c}_j\tau(c_n)}
-\frac{1}{N}(\lambda_j-\bar{\lambda}_j)\delta_{c_n\tau^{-1}(\bar{c}_n)}\right)
\; \delta_{\beta, \tau \backslash n}~. \nonumber \eea This leads directly to
Eq.~(\ref{eqn:emittors}).

\bibliography{bms-equation}

\end{document}